\documentclass[journal]{IEEEtran}
\usepackage{mathptmx}       
\usepackage{helvet}         
\usepackage{courier}        
\usepackage{type1cm}        
\usepackage{makeidx}         
\usepackage{graphicx}        
         
\usepackage{multicol}        
\usepackage[bottom]{footmisc}
\usepackage{ifthen}
\usepackage{subfigure}
\usepackage[usenames,dvipsnames]{color}
\usepackage{cite}

\makeindex            
\usepackage{graphics} 
\usepackage{graphicx} 
\usepackage{epsfig} 
\usepackage{mathptmx} 
\usepackage{times} 
\usepackage{mathtools}  
\usepackage{amssymb}  
\usepackage{amsthm}
\usepackage{amsfonts}
\usepackage{dsfont}
\DeclareGraphicsExtensions{.pdf,.png,.jpg,.eps}

\usepackage{subfig}
\usepackage{verbatim}
\usepackage{comment}
\usepackage{algorithm}
\usepackage{algorithmic}
\usepackage{multirow}
\usepackage{enumitem}
\usepackage{caption}
\usepackage{setspace}
\DeclareGraphicsExtensions{.pdf,.png,.jpg,.eps}

\newcommand{\rr}{{\mathbb{R}}}

\newboolean{showcomments}
\setboolean{showcomments}{true}
\newcommand{\wang}[1]{\ifthenelse{\boolean{showcomments}}
{ \textcolor{red}{(ZW:  #1)}}{}}
\newcommand{\fliu}[1]{\ifthenelse{\boolean{showcomments}}
{ \textcolor{red}{(FL:  #1)}}{}}
\newcommand{\peng}[1]{\ifthenelse{\boolean{showcomments}}
	{ \textcolor{red}{(PY:  #1)}}{}}

\theoremstyle{definition}
\newtheorem{theorem}{Theorem}
\newtheorem{lemma}[theorem]{Lemma}

\newtheorem{proposition}[theorem]{Proposition}

\theoremstyle{definition}
\newtheorem{definition}{Definition}
\newtheorem{remark}{Remark}

\newtheorem{assumption}{Assumption}

\newtheorem{condition}{Condition}

\begin{document}
\setstretch{1}	

\title{Distribute Stability Conditions for Power Systems with Heterogeneous Nonlinear Bus Dynamics}

\author{Peng~Yang, Feng~Liu, \IEEEmembership{Senior~Member,~IEEE,}	
	Zhaojian~Wang,
Chen~Shen, \IEEEmembership{Senior~Member,~IEEE}	
	\thanks{P. Yang, Z. Wang, F. Liu, and C. Shen are with the Department of Electrical Engineering, Tsinghua University, Beijing, 100084, China {\tt\small lfeng@tsinghua.edu.cn}} 
}
\maketitle

\begin{abstract}
	This paper derives distributed conditions that guarantee the system-wide stability for power systems with nonlinear and heterogeneous bus dynamics interconnected via power network. Our conditions require each bus dynamics should satisfy certain passivity-like conditions with a large enough passivity index, a sufficient requirement of which is dictated by the steady-state power flow. The passivity indices uniformly quantify the impacts on the system-wide stability of individual bus dynamics and the coupling strength from the power network. Furthermore, taking three typical bus dynamics as examples, we show that these conditions can be easily fulfilled via proper control design. Simulations on a rudimentary 3-bus example and the IEEE 39-bus system well verify our results under both small and large disturbances.
\end{abstract}


\IEEEpeerreviewmaketitle

\section{INTRODUCTION}
\label{sec:1}
\IEEEPARstart{S}{tability} is the primary concern of power systems and is usually analyzed by centralized methods, such as time-domain simulation\cite{Stott_Powersystemdynamic_1979}, eigenvalue analysis\cite{Sastry_Hierarchicalstabilityalert_1980}, and direct methods\cite{Chiang_Directstabilityanalysis_1995}, all of which rely on the fact that the entire system can be effectively studied as a whole due to the underlying dynamics of power systems only consists of some large synchronous generators (SGs). However, with the proliferation of renewable and distributed energy technologies, the power systems will no longer be dominated by a few large SGs, but by massive small devices with various dynamical characteristics \cite{wang2018distributed_coping}. In such a power system with massive heterogeneous dynamics,  traditional centralized methods may fail due to the computational burden, communication failures, or even privacy concerns \cite{Wang2019Distributed_I,Wang2019Distributed_II,MONSHIZADEH2019258}. Thus, it is necessary to develop distributed stability analytics which is adaptable to heterogeneity and scalability. In this paper, we aim to derive distributed conditions to ensure the system-wide stability for power systems with heterogeneous nonlinear bus dynamics.

One distributed approach is based on decomposition of Jacobian matrix, such as re-constructing the system-wide Jacobian matrix at each agent\cite{Song_DistributedFrameworkStability_2017}, abstracting the interconnection part\cite{Zhang_OnlineDynamicSecurity_2015}, and regarding the effect of interconnection as disturbance\cite{Ilic_standardsmodelbasedcontrol_2012}. However, this approach is limited to linear dynamics and small-signal stability. Other approaches include using 
linear matrix inequalities\cite{Zhang_TransientStabilityAssessment_2016}, sum-of-square technique and vector Lyapunov functions \cite{Kundu_sumofsquaresapproachstability_2015}. These methods may overlook the structural property of the power network and, again, may suffer from computational burden when there are a huge number of bus dynamics.

A more favorable approach is based on the concept of \emph{passivity} or \emph{dissipativity}. This concept has been one of the cornerstones of nonlinear control theory since the 1970s and is widely used in the study of interconnected dynamical systems\cite{Bao_ProcessControlPassive_2007,vanderSchaft_L2GainPassivityTechniques_2017}. In the literature of power systems, it is usually combined with the port-Hamiltonian system framework\cite{VanDerSchaft_PortHamiltoniansystemsnetwork_2004} to study the problem of stability\cite{Fiaz_portHamiltonianapproachpower_2013a,Caliskan_CompositionalTransientStability_2014} and controller design\cite{8424071,Stegink_unifyingenergybasedapproach_2016,trip2016internal}. However, in these works, the network is either \emph{assumed} to be dissipative\cite{Caliskan_CompositionalTransientStability_2014,Stegink_unifyingenergybasedapproach_2016} or is analyzed centrally\cite{Giusto_TransientStabilizationPower_2006,Fiaz_portHamiltonianapproachpower_2013a}.  
Another useful concept is the \textit{passivity index}, which is adaptable to more general cases for both passive and non-passive systems\cite{Sepulchre_ConstructiveNonlinearControl_1997,Li_ConsensusHeterogeneousMultiAgent_2019,Yang_DistributedStabilityAnalytics_2019a}. Passivity index quantifies the excess or shortage of passivity of a dynamical system, which is closely related to stability. And the shortage can be compensated by the excess from other interconnected systems to enforce closed-loop stability\cite{Sepulchre_ConstructiveNonlinearControl_1997}.

In this paper, we tailor the passivity index to the case where the supply rate has differential at the output port, which is referred to as output-differential passivity (OD-passivity). Leveraging this specialized concept, distributed conditions for bus dynamics are derived to ensure the system-wide stability. Once the requirement of minimal passivity index is known, each bus dynamics can locally evaluate whether such conditions are met and adjust its control parameters accordingly.
The main contributions of this paper are twofold: 
\begin{itemize}
	\item The system-wide stability certification is localized on individual bus dynamics, leading to distributed stability conditions that admit heterogeneous nonlinear bus dynamics and further empower scalable stability analytics of power systems. 
	\item A passivity-like index is constructed to quantitatively evaluate the impact of individuals and the coupling strength on the system-wide stability, which can further guide scalable controller designs.
\end{itemize}
The results are illustrated and verified by simulations on a  3-bus example and the IEEE 39-bus benchmark power systems.

The rest of the paper is organized as follows. Section II introduces preliminaries and formulations. The distributed conditions are presented and proved in Section III. In Section IV, we study control designs for three typical bus dynamics to fulfill the proposed conditions. Section V verifies our results through simulations on the 3-bus and IEEE 39-bus power systems.
 Finally, Section VI concludes the paper.

\emph{Notations}: $\rr$ ($\rr_{>0}$) is the set of (positive) real numbers; $\text{col}(x_1,x_2)=(x_1^T,x_2^T)^T$ is a column vector with entries $x_1$ and $x_2$. For a vector $x\in\rr^n$, $\text{diag}\{x\}$ denotes the diagonal matrix with entries from $x$. For a square matrix $A$, let $\lambda_{\min}(A)$ denote its minimal eigenvalue. For a vector $x\in\rr^n$ and a positive semi-definite matrix $M\in\rr^{n\times n}$, define $\|x\|_M:=(x^TMx)^{1/2}$. If $M=I$, it is the Euclidean norm and is simply denoted by $\|x\|$. For a positive number $\delta>0$ and $x^*\in\rr^n$, $\mathcal{U}_\delta(x^*):=\{x\in\rr^n:\;\|x-x^*\|<\delta\}$ denotes the $\delta$-neighborhood of $x^*$. For a symmetric  matrix $A\in\rr^{n\times n}$, $A>(\geq)0$ means $A$ is positive definite (resp. positive semi-definite). Given another symmetric  matrix $B\in\rr^{n\times n}$, $A>(\geq)B$ means $A-B>(\geq)0$.
\section{Preliminaries and Formulations}
\subsection{Modeling Heterogeneous Bus Dynamics}
Consider a network-reduced power system composed of $n$ buses and transmission lines connecting them. It can be abstracted as an undirected graph $\cal G=(\cal V,\cal E)$, where $\cal V$ is the set of buses and $\cal E$ is the set of lines. Each bus in $\cal G$ is associated with a phasor voltage $V_i\angle\theta_i$ and a complex power injection $P_i+\sqrt{-1}Q_i$. $V_i\in\rr_{>0}$ is the magnitude, $\theta_i\in\rr$ is the phase angle. $P_i\in\rr$ and $Q_i\in\rr$ is the real and reactive power injection, respectively. 

Each bus  $i\in \cal V$ is attached to a dynamical power device. For adapting to the heterogeneity, we consider a generic nonlinear input-output model for bus dynamics: 

\begin{equation}
\setlength{\abovedisplayskip}{4pt}	
\setlength{\belowdisplayskip}{4pt}
\label{eq:bus}
\left\lbrace \begin{aligned}
\dot{x}_i&=f_{i}(x_i,u_i)\\
y_i&=C_ix_i
\end{aligned}\right. \;\;\;i\in\mathcal{V}
\end{equation}
where $x_i=\text{col}(\xi_i,\theta_i,V_i)\in\rr^{n_i}\times\rr\times\rr_{>0}$ is the state variable of bus dynamics $i$, and $\xi_i\in\rr^{n_i}$ is the auxiliary state variable which stands for the heterogeneous dynamics of each component. The input is defined as the power output at the bus $u_i=(-P_i,-Q_i/V_i)^T$.\footnote{Here, for the simplicity of notation, we set the second term in $u_i$ as $-Q_i/V_i$ instead of $-Q_i$ directly. Note this formulation does not affect the generality of \eqref{eq:bus} since $f_i(x_i,u_i)$ is a general function of $x_i$ and $u_i$ and hence $Q_i$ can be obtained by $V_i*Q_i/V_i$ for arbitrary $V_i\ne 0$.} The output is selected to be $y_i=C_ix_i=(\theta_i,V_i)^T$.  Let $\mathcal{X}_i:=\rr^{n_i}\times\rr\times\rr_{>0}$.  $f_i:\mathcal{X}_i\times\rr^2\to\mathcal{X}_i$ is a continuously differentiable function.  
  
We generally refer to \eqref{eq:bus}  as bus dynamics since it determines a dynamical relation between the complex voltage and the complex power injection.
The generic model \eqref{eq:bus} covers a broad class of  dynamical models in power systems, such as classical synchronous generators \cite{Stegink_unifyingenergybasedapproach_2016}, inverter-interfaced power devices\cite{Schiffer_surveymodelingmicrogrids_2016,Zhang_OnlineDynamicSecurity_2015,Simpson-Porco_VoltageStabilizationMicrogrids_2017}, and load with frequency/voltage response\cite{Kasis_PrimaryFrequencyRegulation_2017}. Specific examples of  model \eqref{eq:bus} are presented in Section \ref{sec:control}.
\subsection{Modeling the Power Network}
All these sub-systems are interconnected through (lossy) transmission lines, which are represented by the standard admittance matrix $Y=G+jB$. The power flow balance equations at each bus $i\in\mathcal{V}$ are given as follows.
\begin{equation}\label{eq:pf}
\begin{aligned}
P_i&=G_{ii}V_i^2+\sum_{j\in\mathcal{N}_i}V_iV_j(B_{ij}\sin\theta_{ij}+G_{ij}\cos\theta_{ij})\\
Q_i&=-B_{ii}V_i^2-\sum_{j\in\mathcal{N}_i}V_iV_j(B_{ij}\cos\theta_{ij}-G_{ij}\sin\theta_{ij})
\end{aligned}
\end{equation}
where $\mathcal{N}_i$ is the set of nodes who are adjacent to node $i$, $B_{ij}$ and $G_{ij}$ are elements in the admittance matrix.
Let $u:=-\text{col}(P_1,\ldots,P_n,Q_1/V_1,\ldots,Q_n/V_n)$ and $y:=\text{col}(\theta_1,\ldots,\theta_n,V_1,\ldots,V_n)$. The network model \eqref{eq:pf} is a mapping from $y$ to $u$ and can be re-written in a compact form as $u=g(y)$.
\subsection{Modeling the Overall Interconnected System}
\begin{figure}[h]
	\centering
	\setlength{\abovecaptionskip}{0.2cm}	
	\setlength{\belowcaptionskip}{-0.2cm}
	\includegraphics[width=1\hsize]{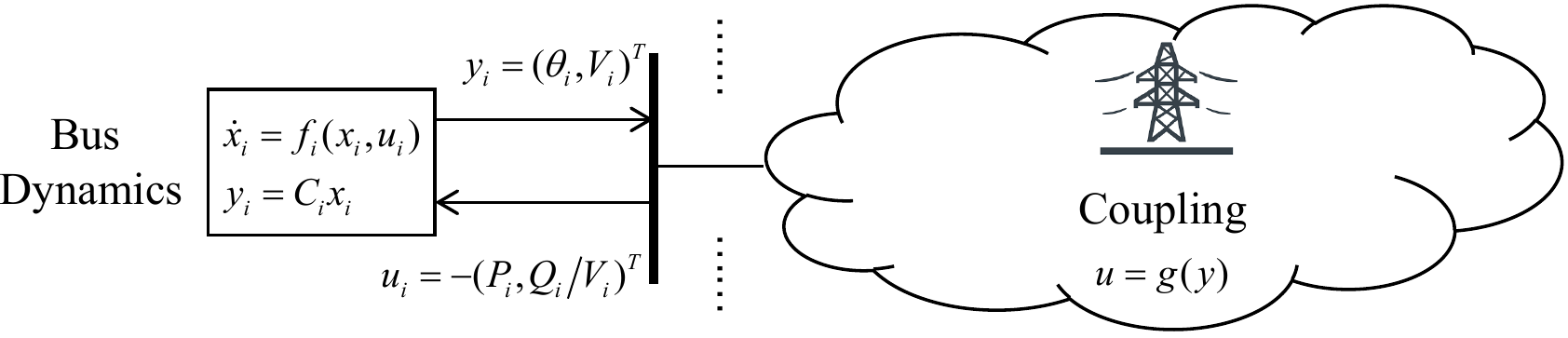}
	\caption{Input-output relation of the bus dynamics and the power network.}
	\label{fig:BusD}
\end{figure}

Combining  \eqref{eq:bus} with  \eqref{eq:pf}, the overall power system can be abstractly formulated in a compact form as follows.
\begin{equation}\label{eq:entire}
\left\lbrace \begin{aligned}
\dot{x}&=f(x,u)\\
y&=Cx\\
u&=g(y)
\end{aligned}\right. 
\end{equation}
The overall system can be further represented as an input-output feedback interconnection between the network and individual bus dynamics as shown in Fig.\ref{fig:BusD}.
\begin{definition}
	$x^*$ is called an equilibrium of system \eqref{eq:entire} if 
	\begin{equation*}
	\left\lbrace \begin{aligned}
	0&=f(x^*,u^*)\\
	y^*&=Cx^*\\
	u^*&=g(y^*)
	\end{aligned}\right. 
	\end{equation*}
	Note that by \eqref{eq:bus} and \eqref{eq:pf} the map from $x^*$ to $u^*,y^*$ is one-to-one. We denote by $(u^*,x^*,y^*)$  the input-state-output triplet associated with the equilibrium $x^*$.
\end{definition}
\begin{assumption}\label{as:isolate}
	The equilibrium $x^*$ exists and is isolated.
\end{assumption}

	Assumption 1  is trivial and commonly used in stability analysis. In the presence of an angle reference, resulting from taking a reference node, the COI coordinate, or an angle-droop control, the equilibrium of the power system is usually isolated or hyperbolic \cite{Kundur_Powersystemstability_1994,chiang1987foundations}.
We aim to construct certain distributed conditions on individual bus dynamics \eqref{eq:bus} such that the equilibrium $x^*$ of the interconnected nonlinear power system \eqref{eq:entire} under Assumption \ref{as:isolate} can be ensured stable.

\subsection{Passivity and Passivity Index}
We now introduce the classical concept of passivity and passivity index, which will be tailored with derivative at the output and employed as the main theoretical tool in this paper. Consider a general input-output nonlinear dynamical system $H: u\mapsto y$
\begin{equation}\label{eq:Sigma}
\left\lbrace \begin{aligned}
\dot{x}&=f(x,u)\\
y&=h(x,u)
\end{aligned}\right. 
\end{equation}
where $x\in \rr^n$, $u\in \rr^m$ and $y\in \rr^m$.

For the simplicity of notations, we assume the zero input-state-output triplet $(u^*,x^*,y^*)=(0,0,0)$. Note,  however, all the following definitions are still valid for nonzero input-state-output triplet $(u^*,x^*,y^*)$ by simply replacing $u$, $x$, and $y$ by $u-u^*$, $x-x^*$, and $y-y^*$.

\begin{definition}[Dissipativity and passivity \cite{willems1972dissipative,Khalil_NonlinearSystems_2002,Sepulchre_ConstructiveNonlinearControl_1997}]\label{de:passive}
	System $H$ is said to be dissipative w.r.t. a supply rate
	$w(t)$  if there exists a non-negative real continuously differentiable function $S:\rr^n\to\rr_{\geq0}$, called a storage function, such that for all $x\in \rr^n$
	and $u\in \rr^m$
	\begin{equation*}
	\setlength{\abovedisplayskip}{4pt}	
	\setlength{\belowdisplayskip}{4pt}
	\dot{S}(x(t))=\frac{\partial S}{\partial x}f(x,u)\leq w(u(t),y(t))
	\end{equation*}
	If $w(t)=u^Ty$, then $H$ is said to be passive.
\end{definition}
\begin{definition}[Output feedback passivity \cite{Khalil_NonlinearSystems_2002,Sepulchre_ConstructiveNonlinearControl_1997}]\label{de:index}
	System 
	$H: u\mapsto y$ is said to be \textit{output feedback passive} (OFP) if it is dissipative w.r.t. the supply rate
	\begin{equation*}
	\setlength{\abovedisplayskip}{4pt}	
	\setlength{\belowdisplayskip}{4pt}
	w(t)=u^Ty-\sigma y^Ty=(u-\sigma y)^Ty
	\end{equation*}
	for some $\sigma\in\rr$, denoted as OFP($\sigma$).
\end{definition}

The real number $\sigma$ quantifies the \textit{excess or shortage} of passivity of system $H$ and is referred to as \emph{passivity index}. A positive $\sigma$ indicates excess of passivity while a negative $\sigma$ indicates shortage \cite{Khalil_NonlinearSystems_2002,Sepulchre_ConstructiveNonlinearControl_1997}. 

Now, we extend these classical definitions to the case where the supply rate has differential at the output port.
\begin{definition}[OD-passivity]\label{de:ODP}
	System 
	$H: u\mapsto y$ is said to be \textit{output-differential passive} (OD-passive) if it is dissipative w.r.t. the supply rate
	\begin{equation*}
	\setlength{\abovedisplayskip}{4pt}	
	\setlength{\belowdisplayskip}{4pt}
	w(t)=u^T\dot{y}-\sigma y^T\dot{y}=(u-\sigma y)^T\dot{y}
	\end{equation*}
	for some $\sigma\in\rr$, denoted as ODP($\sigma$).
\end{definition}
 
Similarly, we say system $H$ has excess (shortage) of OD-passivity if it is ODP($\sigma$) for some $\sigma>0$ ($\sigma<0$).

\begin{definition}[Strict OD-passivity]\label{de:strictODP}
	System 
	$H: u\mapsto y$ is said to be \textit{strictly} OD-passive if it is dissipative w.r.t. the supply rate
	\begin{equation*}
	\setlength{\abovedisplayskip}{4pt}	
	\setlength{\belowdisplayskip}{4pt}
	w(t)=u^T\dot{y}-\sigma y^T\dot{y}-\varphi(\dot{y})
	\end{equation*}
	for some $\sigma\in\rr$, and some positive-definite function $\varphi$ with $\varphi(0)=0$, denoted as SODP($\sigma$).
\end{definition}
Note that, compared with the classical passivity \cite{willems1972dissipative,Khalil_NonlinearSystems_2002,Sepulchre_ConstructiveNonlinearControl_1997}, the supply rate in ODP and SODP is a function of not only $u,y$ but also the differential of output $\dot{y}$. This form of supply rate is related to the Brayton-Moser formulation \cite{Jeltsema_Multidomainmodelingnonlinear_2009a,Yang_DistributedStabilityAnalytics_2019} and the power shaping control technique \cite{Ortega_Powershapingnew_2003,Giusto_TransientStabilizationPower_2006}. Based on the classical notion of passivity, many variations have been proposed to resolve specific problems including supply rates with differentials at the input\cite{Kosaraju_KrasovskiiPassivity_2019} or at the output\cite{Giusto_TransientStabilizationPower_2006}. In Definition \ref{de:strictODP}, strict OD-passivity further requires the dissipation inequality should strictly hold for non-zero $\dot{y}$. This is similar to the classical definition of strictly output passivity (e.g. \cite[Chapter 6]{Khalil_NonlinearSystems_2002}), in which a positive-definite function of output $y$ is employed.

In the next section, we will use these extended passivity concepts to derive distributed stability conditions for power systems with heterogeneous nonlinear bus dynamics.

\section{Distributed Stability Conditions}
\label{sec:4}
The interconnected power system can be divided into two parts: individual bus dynamics, and the electric coupling strength among them. 
Motivated by the concept of passivity index, we quantify the shortage of passivity in the coupling, based on which distributed conditions are derived for individual bus dynamics to ensure the system-wide stability. 
\subsection{Distributed Conditions for Bus Dynamics}
We present distributed conditions for bus dynamics in this subsection, which rely on the input-output property rather than specific dynamical models, such that the heterogeneity of bus dynamics can be dealt with in a unified framework. 

Given an equilibrium $x^*$ of the power system \eqref{eq:entire} and the corresponding input-state-output triplet $(u^*,x^*,y^*)$. For each bus dynamics, consider the following conditions.

\begin{condition}[OD-passivity]\label{c1}
	Each bus dynamics \eqref{eq:bus} in the power system \eqref{eq:entire} is ODP($\sigma_i$) for certain $\sigma_i>0$, in terms of incremental input $\tilde{u}_i:=u_i-u_i^*$ and output $\tilde{y}_i:=y_i-y_i^*$. That is, for each $i\in\mathcal{V}$, there exists a continuously differentiable function $S_i(x_i)$, called \textit{storage function}, such that $x_i^*$ is a local minimum of $S_i$
	and
	\begin{equation}\label{eq:c1}
	\setlength{\abovedisplayskip}{4pt}	
	\setlength{\belowdisplayskip}{4pt}
	\dot{S}_i\leq-(P_i-P_i^*)\dot{\theta}_i-(\frac{Q_i}{V_i}-\frac{Q_i^*}{V_i^*})\dot{V}_i-\sigma_i(y_i-y_i^*)^T\dot{y}_i
	\end{equation}
\end{condition}
\begin{condition}[Strict OD-passivity]\label{c1'}
	Each bus dynamics \eqref{eq:bus} in the power system \eqref{eq:entire} is SODP($\sigma_i$) for certain $\sigma_i>0$, in terms of incremental input $\tilde{u}_i:=u_i-u_i^*$ and output $\tilde{y}_i:=y_i-y_i^*$. That is, for each $i\in\mathcal{V}$, there exists a continuously differentiable function $S_i(x_i)$, called \textit{storage function}, such that $x_i^*$ is a local minimum of $S_i$
	and
	\begin{equation}\label{eq:c1'}
	\setlength{\abovedisplayskip}{4pt}	
	\setlength{\belowdisplayskip}{4pt}
	\dot{S}_i\leq-(P_i-P_i^*)\dot{\theta}_i-(\frac{Q_i}{V_i}-\frac{Q_i^*}{V_i^*})\dot{V}_i-\sigma_i(y_i-y_i^*)^T\dot{y}_i-\varphi_i(\dot{y}_i)
	\end{equation}
	where $\varphi_i$ is a positive-definite function with $\varphi_i(0)=0$.
\end{condition}
\begin{remark}
	Condition \ref{c1} requires each bus dynamics is dissipative w.r.t. the following supply rate
	\begin{equation*}
	\setlength{\abovedisplayskip}{4pt}	
	\setlength{\belowdisplayskip}{4pt}
	w(t)=-(P_i-P_i^*)\dot{\theta}_i-(\frac{Q_i}{V_i}-\frac{Q_i^*}{V_i^*})\dot{V}_i
	-\sigma_i(y_i-y_i^*)^T\dot{y}_i
	\end{equation*}	
	The first two terms $-(P_i-P_i^*)\dot{\theta}_i-(\frac{Q_i}{V_i}-\frac{Q_i^*}{V_i^*})\dot{V}_i$ have a long history in power system community. It appeared in the energy functions \cite{chiang1987foundations,chiang2011direct,Moon_Developmentenergyfunction_1997}, and was explained as the transient energy flow \cite{Ying_energybasedmethodologylocating_2012,chen2013energy}. The third term $-\sigma_i(y_i-y_i^*)^T\dot{y}_i$ is related to the passivity index which we tailored with a  differential at the output. As will be shown later, the index $\sigma_i$ quantifies the impact of each bus dynamics on the system-wide stability. And we will prove that when each individual has enough excess OD-passivity to compensate the non-passivity in the coupling, the stability is ensured.
	
	Condition \ref{c1'} further requires the inequality to be strict for non-zero $\dot{y}_i$, which is indeed technically necessary to derive the asymptotic convergence in Theorem \ref{th:1} and \ref{th:2}. 
\end{remark}

\begin{condition}[Steady-state observability]\label{c2}
	For each bus dynamics \eqref{eq:bus}, a steady output implies a steady state, i.e.
	\begin{equation*}
	\setlength{\abovedisplayskip}{4pt}	
	\setlength{\belowdisplayskip}{4pt}
	\dot{y}_i=0,\;\forall t\geq0\implies \dot{x}_i=0,\;\forall t\geq0
	\end{equation*}
\end{condition}

\begin{remark}
	Condition \ref{c2} can be viewed as a relaxed variation of the classical concept of zero-state observability \cite{Khalil_NonlinearSystems_2002}. The distinction is that only a steady state, i.e. $\dot{x}=0$, is implied in the former, while the specific state $x=0$ is implied in the latter.
	In the context of power systems, this condition means when the voltage $V_i$ and the phase angle $\theta_{i}$ of bus dynamics remain constant, the whole state variable $x_i$ should remain constant as well. Physically, this requires the power device should be stable when it is connected to an ideal voltage source, which holds a constant $V_i\angle\theta_{i}$. And this is widely satisfied for power system devices in practice. 
\end{remark}

We emphasize that Condition \ref{c1}-\ref{c2} are built on the input-output property rather than the detailed model, which enables us to deal with the heterogeneity. And it will be shown in Section \ref{sec:control} that such conditions can be fulfilled by very simple existing controllers.

\subsection{System-Wide Stability: Non-Passivity in The Coupling v.s. Passivity in Bus Dynamics}
The coupling strength plays an important role in the stability of interconnected systems. We will show in this subsection that the system-wide stability can boil down to the competition between non-passivity (shortage of OD-passivity) in the coupling and the (excess) OD-passivity in bus dynamics. 

In the context of power systems, the electric coupling is  depicted by  nonlinear power flow equations \eqref{eq:pf}. It is noticeable that \eqref{eq:pf} consists of $B_{ij}$-related terms and $G_{ij}$-related terms. To streamline the presentation of our results, we first consider the lossless case, i.e. $G_{ij}=0$.
\subsubsection{\textbf{Lossless power system}}
Given $G_{ij}=0$, it can be shown that the gradient of the following function $\tilde{W}_B:\rr^{n}\times\rr_{>0}^n\to\rr$
\begin{equation}\label{eq:WB}
\setlength{\abovedisplayskip}{4pt}	
\setlength{\belowdisplayskip}{4pt}
\tilde{W}_B(y):=\sum_{i\in\mathcal{V}}-\frac{1}{2}B_{ii}V_i^2-\sum_{(i,j)\in\mathcal{E}}B_{ij}V_iV_j\cos\theta_{ij}
\end{equation}
corresponds to the power flow $P$ and $Q/V$, i.e.
\begin{equation}\label{eq:pfB}
\setlength{\abovedisplayskip}{4pt}	
\setlength{\belowdisplayskip}{4pt}
\begin{aligned}
\frac{\partial \tilde{W}_B}{\partial \theta_{i}}&=\sum_{j\in\mathcal{N}_i}B_{ij}V_iV_j\sin\theta_{ij}=P_i\\
\frac{\partial \tilde{W}_B}{\partial V_{i}}&=-B_{ii}V_i-\sum_{j\in\mathcal{N}_i}B_{ij}V_j\cos\theta_{ij}=\frac{Q_i}{V_i}
\end{aligned}
\end{equation}
where the second equality in each sub-equation holds due to $G_{ij}=0$. Note that $\tilde{W}_B(y)$ has a long history in energy functions of power systems \cite{chiang1987foundations,Moon_Developmentenergyfunction_1997,Chiang_Directstabilityanalysis_1995,chiang2011direct}.

Given an equilibrium triplet $(u^*,x^*,y^*)$, we claim that the non-passivity in the coupling of power system \eqref{eq:entire} is quantified by the smallest eigenvalue of the matrix as follows.
\begin{equation}\label{eq:lambda}
\setlength{\abovedisplayskip}{4pt}	
\setlength{\belowdisplayskip}{4pt}
\lambda:=\lambda_{\min}\left(\nabla^2\tilde{W}_B(y^*)\right)
\end{equation}
where $\nabla^2\tilde{W}_B$ is the Hessian matrix of $\tilde{W}_B(y)$. 

Our main theoretical result is that if the OD-passivity of each bus dynamics is enough to compensate the non-passivity in the coupling, 
i.e. the index in Condition \ref{c1} and \ref{c1'} satisfies $\sigma_i>-\lambda,\;\forall i\in\mathcal{V}$,
the system-wide stability of \eqref{eq:entire} is guaranteed.

To streamline the presentation of the stability argument we introduce the following function. For any equilibrium triplet $(u^*,x^*,y^*)$ of \eqref{eq:entire}, consider the network storage function
\begin{equation}\label{eq:SN_loss}
\setlength{\abovedisplayskip}{4pt}	
\setlength{\belowdisplayskip}{4pt}
S_N(y):=W_B(y)+\frac{1}{2}\|y-y^*\|_{\Sigma}^2
\end{equation}
where $\Sigma:=\text{diag}\{(\sigma_1,\ldots,\sigma_n)\}$, and 
\begin{equation}\label{eq:WB2}
	\setlength{\abovedisplayskip}{4pt}	
	\setlength{\belowdisplayskip}{4pt}
	W_B(y):=\tilde{W}_B(y)-(y-y^*)^T\nabla \tilde{W}_B(y^*)-\tilde{W}_B(y^*)
\end{equation}

\begin{lemma}\label{le:1}
	Consider $\lambda$ defined in \eqref{eq:lambda}. Let $\sigma:=\min_{i\in\mathcal{V}}\{\sigma_i\}$. If $\sigma>-\lambda$, then there exists $\delta>0$, such that the network storage function $S_N$ satisfies $S_N(y^*)=0$ and $S_N(y)>0,\,\forall y\in \mathcal{U}_\delta(y^*)\setminus\{y^*\}$.
\end{lemma}

The proof can be found in Appendix \ref{app:A0}.  Lemma \ref{le:1} indicates the minimal requirement of $\sigma$ such that $S_N$ is positive-definite in a certain domain, which will be used later to ensure the system-wide stability. An interesting observation is that a greater $\sigma$ can result in a larger $\mathcal{U}_\delta(y^*)$ in which $S_N(y)$ is positive-definite. This property relates $\sigma$ to the stability region, as discussed in Section \ref{sec:discuss}.
 
Now we are ready to present the main theoretical result.
\begin{theorem}[Lossless power network]\label{th:1}
	Consider a  power system \eqref{eq:entire} with a lossless network. Assume its  equilibrium $x^*$  satisfies Assumption \ref{as:isolate}, and $\lambda$ is defined as \eqref{eq:lambda}. If each bus dynamics $i\in\mathcal{V}$ satisfies 
	\begin{enumerate}
		\item Condition \ref{c1} with $\sigma_i>-\lambda$, then $x^*$ is Lyapunov stable.
		\item Condition \ref{c1'} with $\sigma_i>-\lambda$ and Condition \ref{c2}, then $x^*$ is asymptotically stable.
	\end{enumerate}
\end{theorem}

The proof can be found in Appendix \ref{app:A}.
As a function of $y^*$, $\lambda$ varies with the network operation points, which quantifies the impact of the coupling strength on the system-wide stability. Our simulation results suggest that $\lambda<0$ in most cases, which implies non-passivity and a negative influence, while we may have $\lambda>0$ when the power flow is light. It is also suggested that $\lambda$ will decrease as loads grow and power delivery becomes heavy. That means the stability margin shrinks when the system load is getting heavier, which is consistent with the practice.

	Theorem \ref{th:1} bridges the gap between component-wide OD-passivity and system-wide stability.  It is implied that when the excess of OD-passivity in bus dynamics is sufficient to compensate the non-passivity in the coupling, i.e. $\sigma_i+\lambda>0$, the stability of the  interconnected system can be ensured. 
\subsubsection{\textbf{Lossy power system}} Now we consider the lossy power system with non-zero $G_{ij}$. The power flow related to the transfer conductance $G_{ij}$ is given as follows.
\begin{equation}\label{eq:pfG}
\setlength{\abovedisplayskip}{4pt}	
\setlength{\belowdisplayskip}{4pt}
\begin{aligned}
\phi_{pi}(y)&:=G_{ii}V_i^2+\sum_{j\in\mathcal{N}_i}G_{ij}V_iV_j\cos\theta_{ij}\\
\phi_{qi}(y)&:=\sum_{j\in\mathcal{N}_i}G_{ij}V_j\sin\theta_{ij}
\end{aligned}
\end{equation}
Let $\Phi(y):=\text{col}(\phi_{p1},\ldots,\phi_{pn},\phi_{q1},\ldots,\phi_{qn})$.
Consider again the function $\tilde{W}_B(y)$ defined in \eqref{eq:WB}. For lossy power system we have for all $i\in\mathcal{V}$
\begin{equation}\label{eq:pfsum}
\setlength{\abovedisplayskip}{4pt}	
\setlength{\belowdisplayskip}{4pt}
P_i=\frac{\partial \tilde{W}_B}{\partial \theta_{i}}+\phi_{pi},\qquad\frac{Q_i}{V_i}=\frac{\partial \tilde{W}_B}{\partial V_{i}}+\phi_{qi}
\end{equation}
Unlike $B_{ij}$-related terms, the $G_{ij}$-related part has always been an obstacle to deriving Lyapunov function or energy function for the lossy power system. It can be shown that $\Phi(y)$ is a non-conservative vector field and thus causes path-dependent integral, which hinders the construction of a well-defined Lyapunov function \cite{Chiang_Directstabilityanalysis_1995}. Although extensive efforts have been put into this challenging issue, unfortunately, no successful solution is obtained yet. Alternatively, the so-called \textit{numerical} energy function is widely adopted in the power system community\cite{Chiang_Directstabilityanalysis_1995,athay1979practical}, which is essentially an approximation of $G_{ij}$-related term. Despite the theoretical imperfection, this kind of function has been shown to be a highly acceptable approximation with satisfactory performances in both simulations and industrial practices \cite{chiang2013line}.

In this paper, we follow the lines of numerical energy functions to deduce the stability result for lossy power systems. 
For any given equilibrium triplet $(u^*,x^*,y^*)$, consider the quadratic function as follows.
\begin{equation}\label{eq:WG}
\setlength{\abovedisplayskip}{4pt}	
\setlength{\belowdisplayskip}{4pt}
W_G(y):=\frac{1}{2}(y-y^*)^T\nabla\Phi(y^*)(y-y^*)
\end{equation}
Taylor expansion yields
\begin{equation}
\setlength{\abovedisplayskip}{4pt}	
\setlength{\belowdisplayskip}{4pt}
\Phi(y)=\Phi(y^*)+\nabla\Phi(y^*)(y-y^*)+O(\|y-y^*\|^2)
\end{equation} 
Omitting the higher-order infinitesimal leads to 
\begin{equation}\label{eq:approx}
\setlength{\abovedisplayskip}{4pt}	
\setlength{\belowdisplayskip}{4pt}
\nabla W_G(y)\approx\Phi(y)-\Phi(y^*)
\end{equation}
In this way, the path-dependency of $\Phi(y)$ is avoided. Such an approximation can be regarded as a variant of the linear-path approximation used in \cite{Chiang_Directstabilityanalysis_1995} and \cite{athay1979practical}.

In this context, we claim that the non-passivity in the coupling of power system \eqref{eq:entire} can be quantified by the smallest eigenvalue of the matrix as follows.
\begin{equation}\label{eq:lambda2}
\setlength{\abovedisplayskip}{4pt}	
\setlength{\belowdisplayskip}{4pt}
{\lambda}:=\lambda_{\min}\left(\nabla^2W_B(y^*)+\frac{1}{2}\left(\nabla\Phi(y^*)+\nabla\Phi(y^*)^T\right)\right)
\end{equation}
Similar to Theorem \ref{th:1}, the system-wide stability can be ensured if the OD-passivity in individual bus dynamics is sufficient to compensate all the non-passivity in the coupling, which in this case is quantified by $\lambda$ in \eqref{eq:lambda2} instead of \eqref{eq:lambda}.
\begin{theorem}[Lossy power network]\label{th:2}
	Consider a lossy power system \eqref{eq:entire} with approximation \eqref{eq:approx}. Assume its equilibrium $x^*$  satisfies Assumption \ref{as:isolate}, and  $\lambda$  is defined as \eqref{eq:lambda2}. If each bus dynamics $i\in\mathcal{V}$ satisfies 
	\begin{enumerate}
		\item Condition \ref{c1} with $\sigma_i>-{\lambda}$, then $x^*$ is Lyapunov stable.
		\item Condition \ref{c1'} with $\sigma_i>-{\lambda}$ and Condition \ref{c2}, then $x^*$ is asymptotically stable.
	\end{enumerate}
\end{theorem}

	The proof can be found in Appendix \ref{app:B}.
\begin{remark}
Theorem \ref{th:2} reveals that network loss essentially affects the non-passivity in the electric coupling and may undermine the system-wide stability as a consequence. In this case, more OD-passivity should be provided by bus dynamics to ensure stability. 
While Theorem \ref{th:2} is established based on the approximation \eqref{eq:approx}, 
our  simulations empirically justify its validation, which will be illustrated  in Section \ref{sec:case}. 
\end{remark}

\subsection{Discussion on the Stability Region}\label{sec:discuss}
Theorems \ref{th:1} and \ref{th:2}  claim that $\sigma_i>-\lambda$, combined with other conditions, leads to the asymptotic stability. Then another question arises naturally: what if an even bigger $\sigma_i$ is provided?
Here we provide a simple discussion by considering the relation between $\sigma_i$ and the stability region.

Assume a power system \eqref{eq:entire} with all individual bus dynamics satisfying Conditions \ref{c1}-\ref{c2}. In the proofs of Theorems \ref{th:1} and \ref{th:2}, the function $W(x)$ defined as \eqref{eq:Lyapunov} is verified as a Lyapunov function in the domain $D$. Thus, the largest bounded level set $\Omega_c:=\{x:W(x)\leq c\}$ contained in $D$ could serve as an estimation of the stability region. It follows from the proofs of Lemma \ref{le:1} and Theorems \ref{th:1} and \ref{th:2} that fixing other factors, a bigger $\sigma_i$ results in a larger domain $D$, in which $W(x)$ is a valid Lyapunov function. 

Based on this observation, we suggest, that without proof but backed up by simulations, the greater $\sigma_i$ is (within a certain saturation extend), the larger stability region will be.

\section{Fulfill The Conditions via Control}\label{sec:control}
In this section, we aim to show how the proposed conditions can be fulfilled by the existing power devices. Three typical examples of the generic model \eqref{eq:bus} are studied here for the purpose of illustration. It is proven that very simple controllers with proper parameters are enough to equip these devices with desired properties.
\subsection{Synchronous Generators}
A typical power device of model \eqref{eq:bus} is the flux-decay model of SG as follows\cite{Stegink_unifyingenergybasedapproach_2016}.
\begin{equation}\label{eq:sg1}
\setlength{\abovedisplayskip}{4pt}	
\setlength{\belowdisplayskip}{4pt}
\left\lbrace 
\begin{aligned}
\dot{\delta}_i&=\omega_i\\
M_i\dot{\omega}_i&=-D_i\omega_i-P_i+P^g_i\\
T_{di}'\dot{E}_{qi}'&=-E_{qi}'-\frac{x_{di}-x_{di}'}{E_{qi}'}Q_i+E_{fi}
\end{aligned}\right. 
\end{equation}
where $E_{qi}'\angle\delta_i$ is the $q$-axis transient internal voltage. $\omega_i$ is the frequency derivation. $M_i$ is the moment of inertia. $D_i$ is the damping coefficient. $T_{di}'$ is the $q$-axis open-circuit transient time constant. $x_{di}$ and $x_{di}'$ are the $d$-axis synchronous reactance and transient reactance, respectively. For a realistic SG, $x_{di}>x_{di}'$. $P_i^g$ and $E_{fi}$ are control signals, which stand for the power generation and the excitation voltage, respectively. $P_i$ and $Q_i$ are the output active and reactive power, respectively.

\begin{proposition}[Synchronous Generator]\label{pro:sg}
	For any given $\sigma_i\in\rr$, set the control of \eqref{eq:sg1} as
	\begin{subequations}\label{eq:sgcontrol}
		\setlength{\abovedisplayskip}{4pt}	
		\setlength{\belowdisplayskip}{4pt}
		\begin{equation}\label{eq:sgPg}
		P_i^g=-K_I\int_{0}^{\tau}\omega_idt-K_P\omega_i+P_i^{g*}
		\end{equation}
		\begin{equation}\label{eq:sgEf}
		E_{fi}=-K_E(E_{qi}'-E_{qi}'^*)+E_{fi}^*
		\end{equation}
	\end{subequations}
	where $K_I>\sigma_i$, $K_P>-D_i$ and $K_E>(x_{di}-x_{di}')\sigma_i-1$ are constant. $P_i^{g*}$ and $E_{fi}^*$ are the steady-state inputs up to the setting point. Then the bus dynamics \eqref{eq:sg1} satisfies Conditions \ref{c1}-\ref{c2}.
\end{proposition}

	The proof is given in Appendix \ref{app:C}. Note that \eqref{eq:sgPg} is a standard PI controller with the frequency derivation $\omega_i$ as input, and \eqref{eq:sgEf} is simply a negative feedback. This simple and classical controller is enough to meet Conditions \ref{c1}-\ref{c2}.

\subsection{Conventional Droop-Controlled Inverters}
Another typical bus dynamics of model \eqref{eq:bus} is the inverter-interfaced device with conventional $P-\theta$ and $Q-V$ droop control as follows\cite{Zhang_OnlineDynamicSecurity_2015}.
\begin{equation}\label{eq:egV}
\setlength{\abovedisplayskip}{4pt}	
\setlength{\belowdisplayskip}{4pt}
\left\lbrace 
\begin{aligned}
\tau_{i1}\dot{\theta}_i&=-(\theta_i-\theta_i^*)-D_{i1}(P_i-P_i^*)\\
\tau_{i2}\dot{V}_i&=-(V_i-V_i^*)-D_{i2}(Q_i-Q_i^*)
\end{aligned}\right. 
\end{equation}
where $\tau_{i1}$, $\tau_{i2}$ are the time constants and $D_{i1}$, $D_{i2}$ are the droop gains. Note that \eqref{eq:egV} could be either a source or a load depending on the sign of $P_i^*$ and $Q_i^*$.

\begin{proposition}[Conventional Droop]\label{pro:CD}
	For any given $\sigma_i\in\rr$, let the droop gains $D_{i1}\in\rr_{>0}$ and $D_{i2}\in\rr_{>0}$ in \eqref{eq:egV} satisfy
	\begin{equation}\label{eq:conventionD}
	\setlength{\abovedisplayskip}{4pt}	
	\setlength{\belowdisplayskip}{4pt}
	D_{i1}^{-1}>\sigma_i,\quad D_{i2}^{-1}>(V_i^{*2}\sigma_i-Q_i^*)/V_i^*
	\end{equation}
	Then the bus dynamics \eqref{eq:egV} satisfies Conditions \ref{c1}-\ref{c2}.
\end{proposition}

	The proof is given in Appendix \ref{app:D}.

\subsection{Quadratic Droop-Controlled Inverters}
We now consider another typical bus dynamics in the literature, which is known as the quadratic droop controller in the literature \cite{Simpson-Porco_VoltageStabilizationMicrogrids_2017}. Its dynamics can be expressed as follows.
\begin{equation}\label{eq:egQD}
\setlength{\abovedisplayskip}{4pt}	
\setlength{\belowdisplayskip}{4pt}
\left\lbrace 
\begin{aligned}
\tau_{i1}\dot{\theta}_i&=-(\theta_i-\theta_i^*)-D_{i1}(P_i-P_i^*)\\
\tau_{i2}\dot{V}_i&=-D_{i2}Q_i-V_i(V_i-u_i^*)
\end{aligned}\right. 
\end{equation}
where $u_i^*$ is a constant satisfying
\begin{equation}\label{eq:QDu}
\setlength{\abovedisplayskip}{4pt}	
\setlength{\belowdisplayskip}{4pt}
0=-D_{i2}Q_i^*-V_i^*(V_i^*-u_i^*)
\end{equation}

\begin{proposition}[Quatratic Droop]\label{pro:QD}
	For any given $\sigma_i\in\rr$, let the droop gains $D_{i1}\in\rr_{>0}$ and $D_{i2}\in\rr_{>0}$ in \eqref{eq:egQD} satisfy
	\begin{equation}\label{eq:QDroop}
	\setlength{\abovedisplayskip}{4pt}	
	\setlength{\belowdisplayskip}{4pt}
	D_{i1}^{-1}>\sigma_i,\quad D_{i2}^{-1}>\sigma_i
	\end{equation}
	Then the bus dynamics \eqref{eq:egQD} satisfies Conditions \ref{c1}-\ref{c2}.
\end{proposition}

	The proof is given in Appendix \ref{app:E}.

 Proposition \ref{pro:CD} and \ref{pro:QD} both require the droop gains less than certain values, which is consistent with the result that a too large gain can cause instability \cite{Pogaku_ModelingAnalysisTesting_2007b}.

\begin{remark}
	It is worth mentioning that the applicable scope of our method is not limited to these illustrative examples. The proposed Conditions \ref{c1}-\ref{c2} are applicable to general dynamics \eqref{eq:bus}, provided a suitable storage function $S_i$. Since we adopt the general framework of passivity, many well-developed techniques can be exploited to construct $S_i$, e.g. the LMI technique and the SOS method \cite{Khalil_NonlinearSystems_2002,vanderSchaft_L2GainPassivityTechniques_2017}. Note, however, it may not be trivial to find a suitable storage function for the SG with no frequency integral. The stability issue, in that case, falls in a more general topic--the stability of synchronization. One possible way to extend our methods to that more general case is following the line of \cite[Proposition 5.9]{schiffer2014conditions}, which transfers the requirement on $\theta$ to $V$ dynamics.
\end{remark}

\begin{remark}[Implementation issues]
	The system-wide stability is ensured if each bus dynamics properly sets its controllers such that its passivity index $\sigma_i$ is greater than the network-side index $-\lambda$. Since $\lambda$ is only dictated by the steady-state power flow and is independent of specific bus dynamics, it enables a distributed approach to assess the system-wide stability. For practical implementation, there are different ways to realize such a distributed framework. Here, we provide two possible approaches. First, with substantial historical power flow data, which is generally available at dispatch centers, the system operator can estimate the value of $\lambda$ for certain operation situations and set it as an access protocol for power devices in the grid. This approach may be conservative but is free from equilibrium changes. Second, provided a well-developed communication network, the system operator can calculate $\lambda$ online and broadcast it to every connected power device. Then each device can adjust its control parameters accordingly. In this way, the conservativeness can be reduced but certain communication infrastructure is necessary.
\end{remark}

\section{Case Study}\label{sec:case}
\subsection{The 3-Bus Power System}
\begin{figure}[h]
	\centering
	\setlength{\abovecaptionskip}{0.2cm}	
	\setlength{\belowcaptionskip}{-0.2cm}
	\includegraphics[width=0.8\hsize]{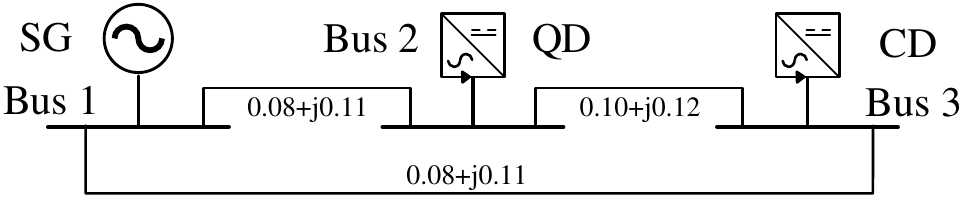}
	\caption{The schematic of the 3-bus power system.}
	\label{fig:3bus}
\end{figure}

For the purpose of illustration, we first give a 3-Bus power system example consisting of a SG, a conventional droop (CD), and a quadratic droop (QD) as shown in Fig.\ref{fig:3bus}. Impedance $r+jx$ of each line in per unite (p.u.) is depicted in Fig.\ref{fig:3bus} as well. Other parameters are listed in Tab. \ref{tab1}.
\begin{table}[h]
	\centering
	\footnotesize
	\caption{Parameters of the 3-bus system.}
	\label{tab1}
	\begin{tabular}{l|l}
		\hline
		Parameters&Values\\
		\hline
		SG: $M_i,D_i,T_{di}',x_{di},x_{di}'$ /p.u.   & 0.16, 0.076, 6.56, 0.295, 0.17 \\ 
		QD: $\tau_{i1},\tau_{i2}$ /s & 0.7, 8 \\ 
		CD: $\tau_{i1},\tau_{i2}$ /s& 1, 10 \\
		\hline
	\end{tabular}
\end{table}

To solve the power flow and obtain the system equilibrium, we set bus 1 as the $V\theta$ node, bus 2 as a $PV$ node, and bus 3 as a $PQ$ node. The base load profile is $P_2=1,P_3=-1.5,Q_3=-0.1$. In order to show the characteristics of $\lambda$ changing with the load, the base profile is multiplied by a scale factor $s$, from 0.5 to 2.5, to simulate different load conditions.

We set the controls of each bus dynamics according to Proposition \ref{pro:sg}-\ref{pro:QD} with a uniform passivity index $\sigma=\sigma_1=\sigma_2=\sigma_3$. And for each scale factor $s$, to verify our theoretical results, the proposed condition $\sigma>-\lambda$ is compared to the exact minimal $\sigma$ for maintaining stability, which is obtained by the system-wide eigenvalue analysis. It should be noted that our theoretical results are based on nonlinear approaches, in which the Lyapunov function is provided and can be further exploited for other nonlinear analyses, e.g. to estimate the stability region \cite{Khalil_NonlinearSystems_2002}. Since the small-signal stability is a necessary condition and a natural corollary of the large-signal stability, we just used Fig. \ref{fig:tight_3bus}(A) (and also following Fig. \ref{fig:nonuniform3bus}, \ref{fig:tight} and \ref{fig:tight2}) as verification of our results. 

The result is displayed in Fig.\ref{fig:tight_3bus}A.
It is shown that $\sigma>-\lambda$ is sufficient for system-wide stability and it seems to be also necessary for this case. It is also noticeable that the more stressful the system is, the more non-passivity shows up in the coupling and the more passivity from each bus dynamics should be injected into the network to ensure system-wide stability.

\begin{figure}[h]
	\centering
	\footnotesize
	\setlength{\abovecaptionskip}{0.2cm}	
	\setlength{\belowcaptionskip}{-0.2cm}
	\includegraphics[width=1\columnwidth]{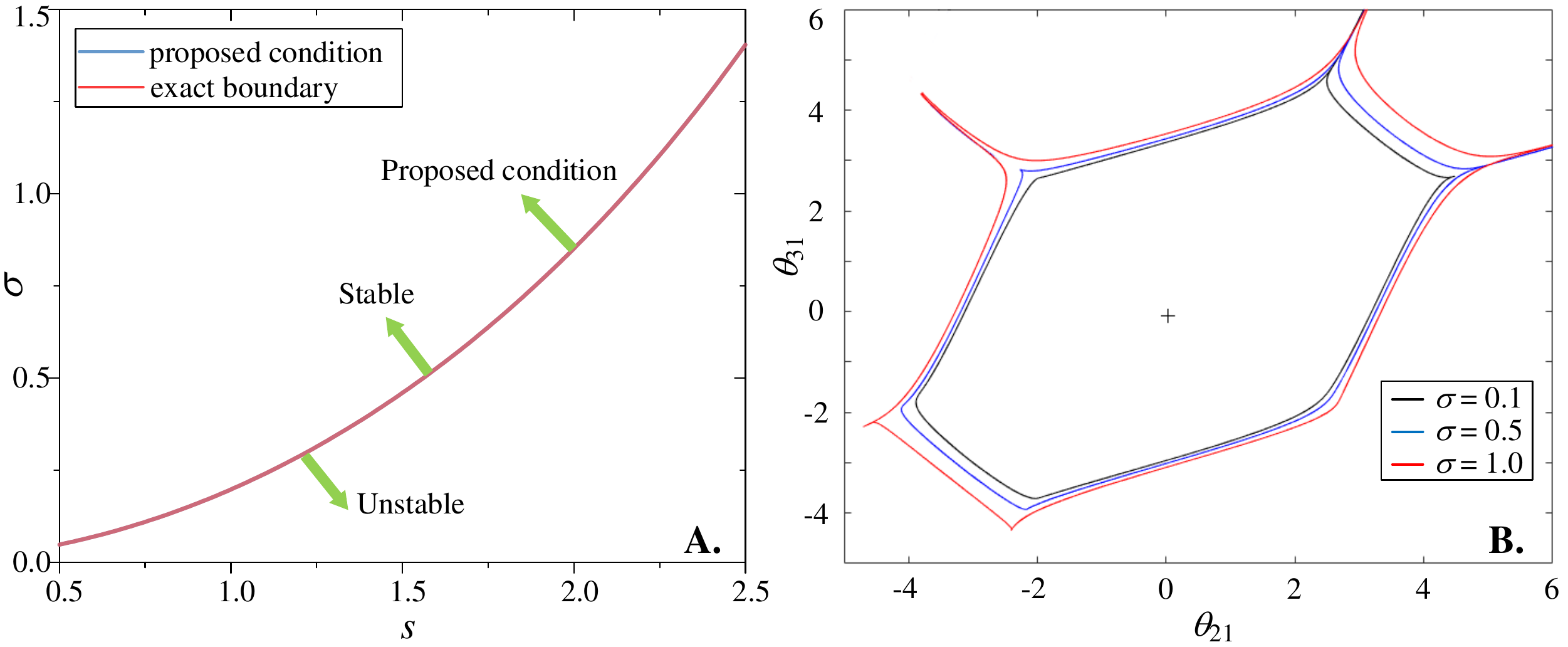}
	\caption{(A) Stability verification of the 3-bus system. The proposed condition and the exact boundary almost overlap in this case. (B) Projection on phase angle plane of stability boundary of the 3-bus system with different $\sigma$.}
	\label{fig:tight_3bus}
\end{figure}
To illustrate the relation between $\sigma$ and the stability region, consider the equilibrium in the base load profile. We first calculate the stability boundary of the original system then project it to the phase angle plane with $\theta_1$ as the reference, on which other state variables except $\theta_i$ are fixed at their steady-state values. The results under different $\sigma$ are depicted in Fig.\ref{fig:tight_3bus}B. It is suggested that greater passivity indices result in larger stability regions, which confirms our conjecture in Section \ref{sec:discuss}.

Now consider each bus dynamics has a non-uniform passivity index. With load scale factor $s=1.5$, our proposed condition requires $\sigma_i>-\lambda=0.4599$ for $i=1,2,3$. Assume devices in bus $\#$1 and $\#$2 take varying passivity indexes while bus $\#$3 fixes its index at $\sigma_3=0.46$. The region of our proposed condition and the boundary for system-wide small-signal stability are shown in Fig. \ref{fig:nonuniform3bus}. This verifies our results as a sufficient condition for system-wide stability. Given all $\sigma_i>-\lambda$, no matter uniform or not, the system-wide stability is always ensured. On the other hand, if some bus dynamics fail to meet the requirement, then the system is possibly (but not necessarily) unstable. 
\begin{figure}[h]
	\centering
	\setlength{\belowcaptionskip}{-0.2cm}
	\includegraphics[width=0.7\columnwidth]{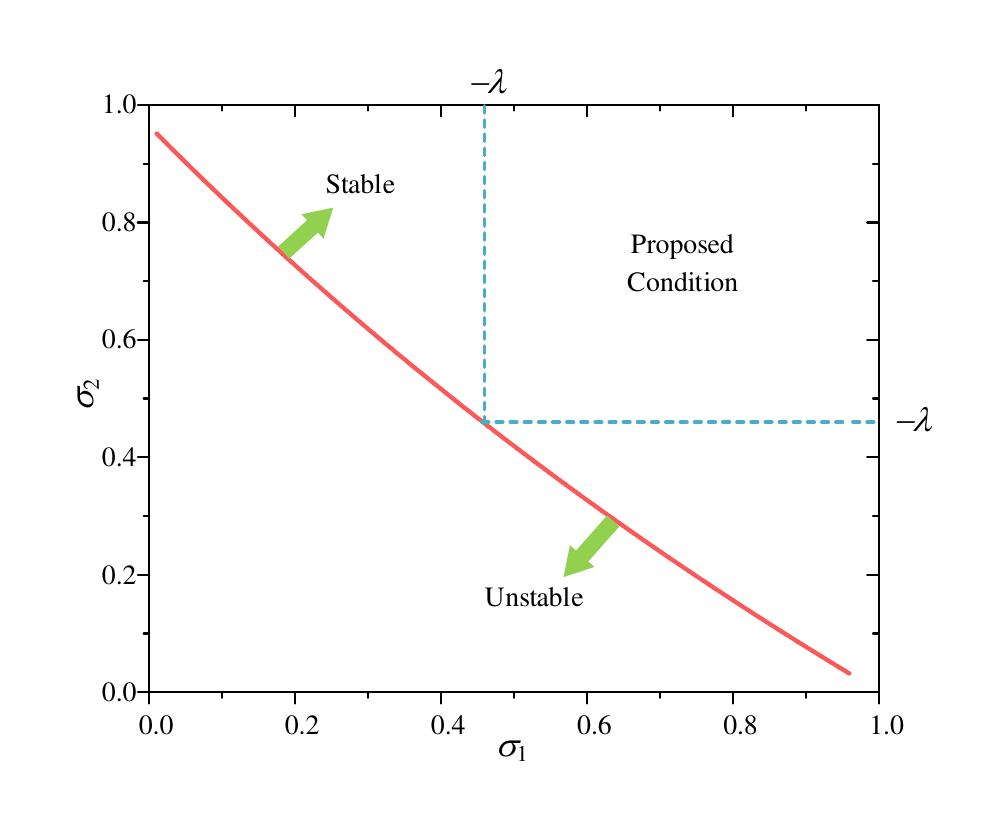}
	\caption{Stability verification of the 3-bus system with non-uniform passivity indexes ($\sigma_3=0.46>-\lambda$).}
	\label{fig:nonuniform3bus}
\end{figure}

The non-uniform case raises an interesting implication about the responsibility for system-wide stabilization. Since the passivity index quantifies the impact of each bus dynamics, it quantifies the responsibility as well. If every participant in the power system should equally share the responsibility for stabilization, it is fair to blame the bus dynamics with $\sigma_i<-\lambda$ for the possible instability, while others are not liable. The equal-share principle seems to be a promising solution for a future power system with massive distributed devices and it the starting point of our study. 
\subsection{The IEEE 39-Bus System}
\begin{figure}[h]
	\centering
	\setlength{\abovecaptionskip}{0.2cm}	
	\setlength{\belowcaptionskip}{-0.2cm}
	\includegraphics[width=0.9\hsize]{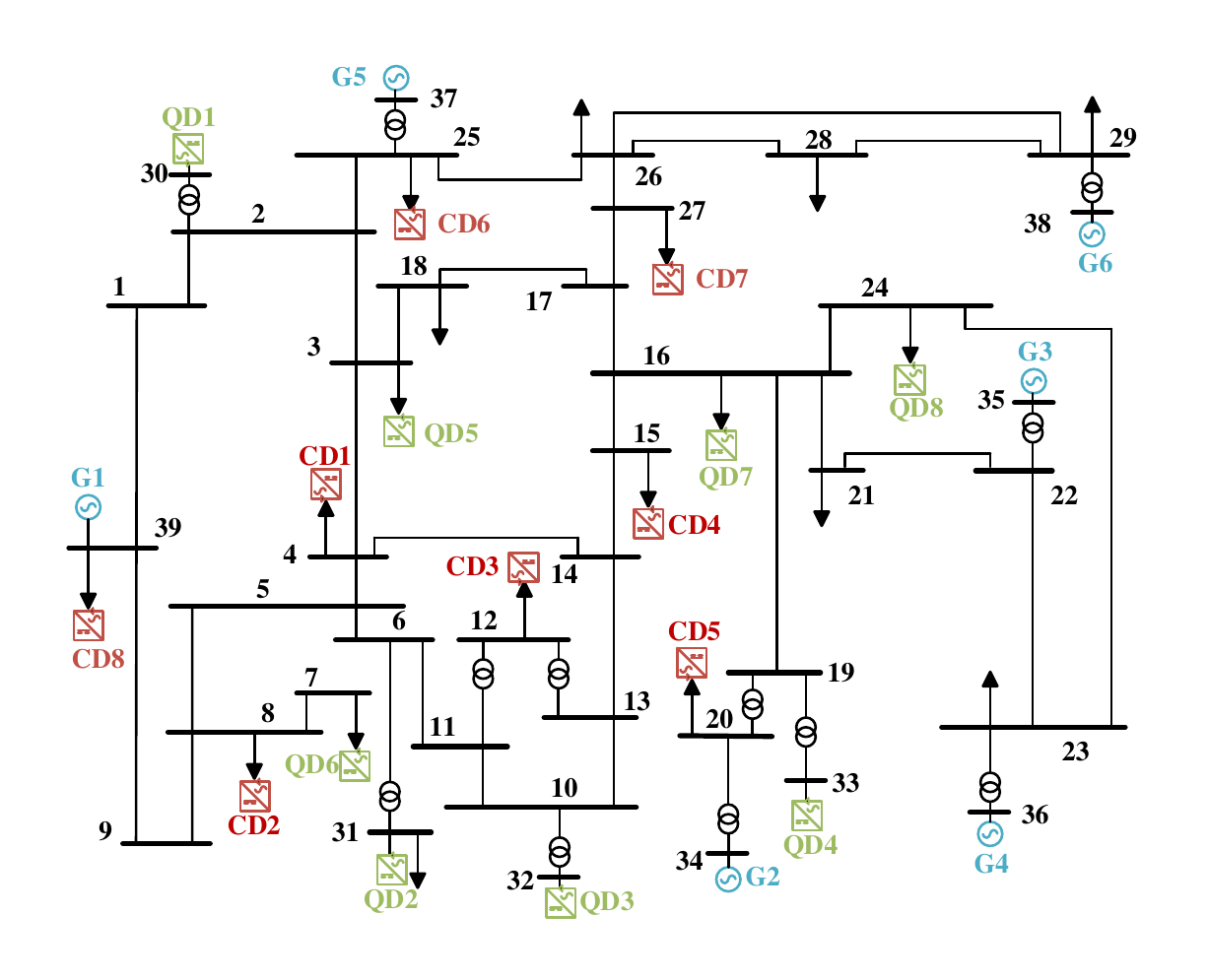}
	\caption{The schematic of the IEEE 39-bus power system.}
	\label{fig:39bus}
\end{figure}

To demonstrate the scalability of our results, now consider the IEEE 39-bus power system benchmark. In our simulation, we alter the original system by substituting four SGs into four QDs and equipping twelve loads with eight CDs and four QDs, as shown in Figure~\ref{fig:39bus}. The parameters of SGs and the power network can be found in \cite{hiskens2013ieee}. The time constants of droops are listed in Tab. \ref{tab2}. There are in total 22 dynamical components that form 50th-order differential equations.

\begin{table}[h]
	\centering
	\footnotesize
	\caption{Time constants (s) in the IEEE 39-bus system.}
	\label{tab2}
	\begin{tabular}{l|llllllll}
		\hline
		&QD1&QD2&QD3&QD4&QD5&QD6&QD7&QD8\\
		\hline
		$\tau_{i1}$ & 0.3&0.2&0.13&0.26&0.4&0.2&0.25&0.35  \\ 
		$\tau_{i2}$ & 7&7.5&8&8.2&8.5&6.5&9.2&9.6\\
		\hline
		&CD1&CD2&CD3&CD4&CD5&CD6&CD7&CD8\\
		\hline
		$\tau_{i1}$&0.3&0.25&0.15&0.28&0.34&0.22&0.4&0.5\\
		$\tau_{i2}$ &8.1&9&9.5&9.3&10&9.5&7&6.5\\
		\hline
	\end{tabular}
\end{table}

\begin{figure}[h]
	\centering
	\setlength{\abovecaptionskip}{0.2cm}	
	\setlength{\belowcaptionskip}{-0.2cm}
	\includegraphics[width=0.7\columnwidth]{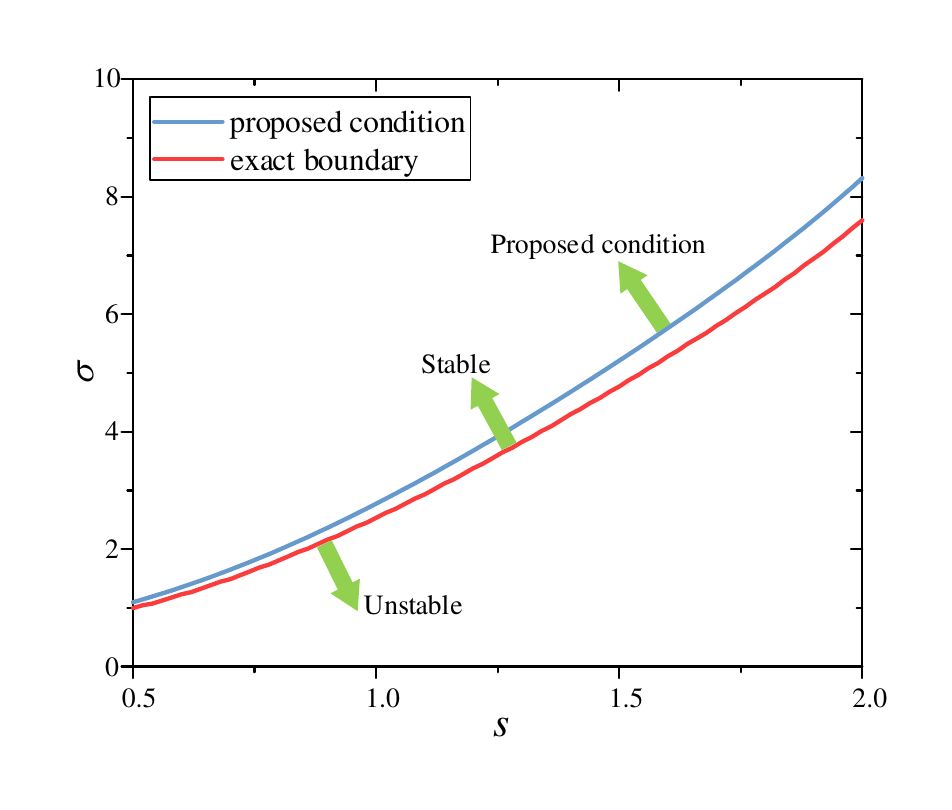}
	\caption{Stability verification of the IEEE 39-bus power system with 22 heterogeneous bus dynamics.}
	\label{fig:tight}
\end{figure}

To show the characteristics of $\lambda$ changing with the load, the original power flow setting in \cite{hiskens2013ieee} is multiplied by a scale factor $s$, from 0.5 to 2, to simulate different load levels. 
We again set the controls of each bus dynamics according to Proposition \ref{pro:sg}-\ref{pro:QD} with a uniform passivity index $\sigma_i=\sigma$, $\forall i\in\mathcal{V}$. And for each scale factor $s$, we calculate the proposed condition $\sigma>-\lambda$ and the exact minimal $\sigma$ for stability obtained by eigenvalue analysis. The results are shown in Fig.\ref{fig:tight}.

It shows that the coupling has a deficit and its index decreases as the loads scale up, which implies that the stability margin deteriorates and bus dynamics should support more passivity to maintain system-wide stability.
It is clear that the system-wide stability is always guaranteed if every bus obeys $\sigma_i>-\lambda$, which justifies our theoretical results. The gap between these two lines indicates the conservativeness of our conditions. It is worth mentioning that the difference between the two curves does not result from the approximation \eqref{eq:approx}, since \eqref{eq:approx} holds strictly if only the small-signal stability is concerned. However, the error caused by the approximation has to be noted when the transient stability is concerned.

\begin{figure}[h]
	\centering
	\setlength{\abovecaptionskip}{0.2cm}	
	\setlength{\belowcaptionskip}{-0.2cm}
	\includegraphics[width=1\columnwidth]{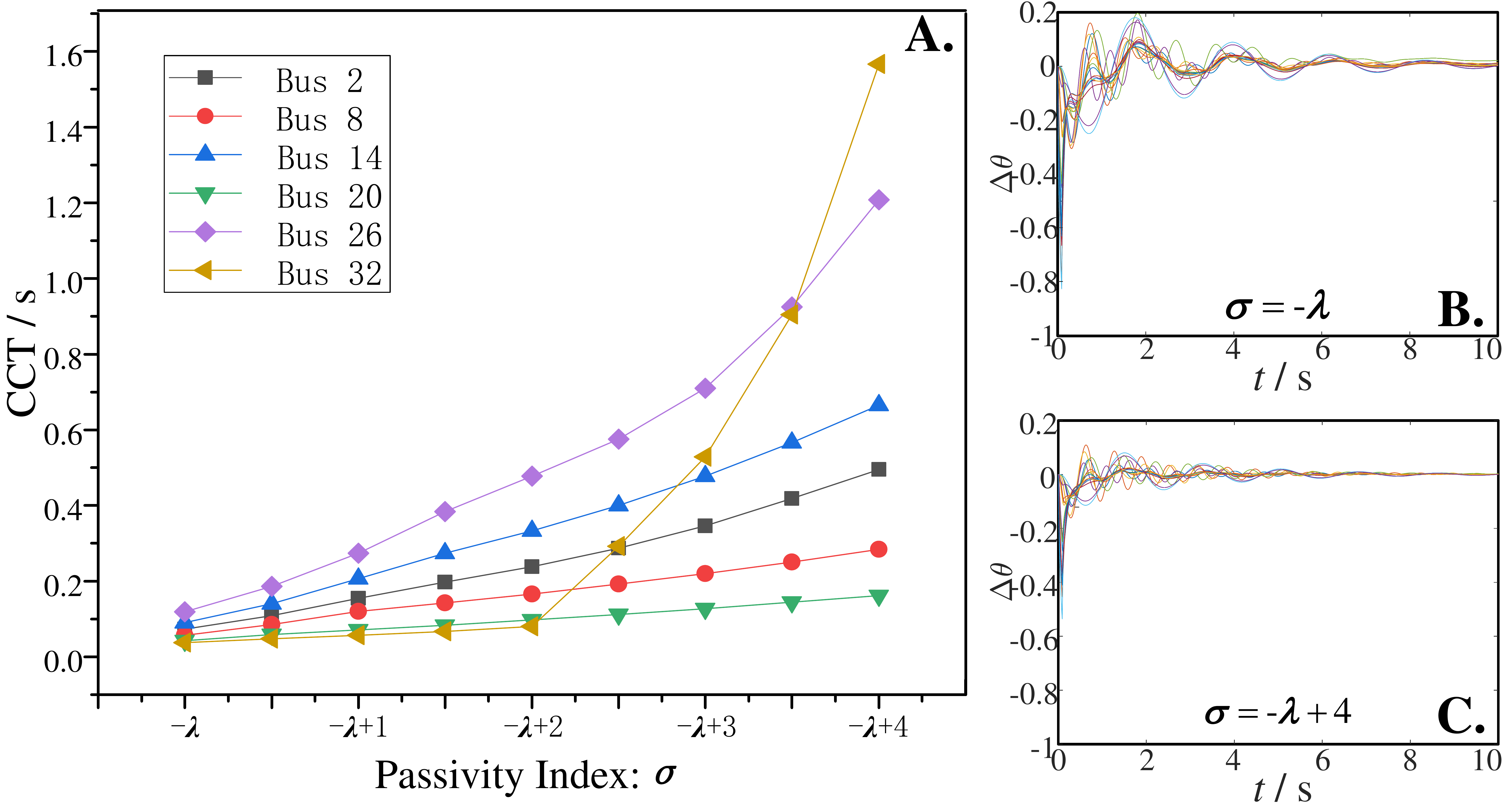}
	\caption{Transient Stability of IEEE 39-bus system with different passivity index $\sigma$. (A) CCTs for three-phase short circuit fault at six different bus increase with $\sigma$; (B) and (C) the trajectories of $\Delta\theta$ for the fault that occurs at bus 14 and is cleared at 0.08s, with $\sigma=-\lambda$ and $\sigma=-\lambda+4$, respectively.}
	\label{fig:CCT}
\end{figure}

Now we illustrate the relation between the passivity index and the system-wide transient stability. Consider the base load profile, i.e. $s=1$. A three-phase short circuit fault occurs at six different buses, respectively. For each fault, we vary the uniform passivity index $\sigma$ from $-\lambda$ to $-\lambda+4$. For each case, the critical clearing time (CCT) is calculated via detailed numerical simulation. The results of CCTs are shown in Fig.\ref{fig:CCT}, as well as the trajectories of angle derivation for fault at bus 14 with $\sigma=-\lambda$ and $\sigma=-\lambda+4$, respectively.

Fig.\ref{fig:CCT}A shows that CCTs increase with $\sigma$ for all faults, which implies a larger stability region of the power system. Also, comparing Fig.\ref{fig:CCT}B and C shows that a greater $\sigma$ results in better transient response. These observations suggest that the more passivity provided by bus dynamics, the more transient stable the system is.

\begin{figure}[h]
	\centering
	\setlength{\abovecaptionskip}{0.2cm}	
	\setlength{\belowcaptionskip}{-0.2cm}
	\includegraphics[width=1\columnwidth]{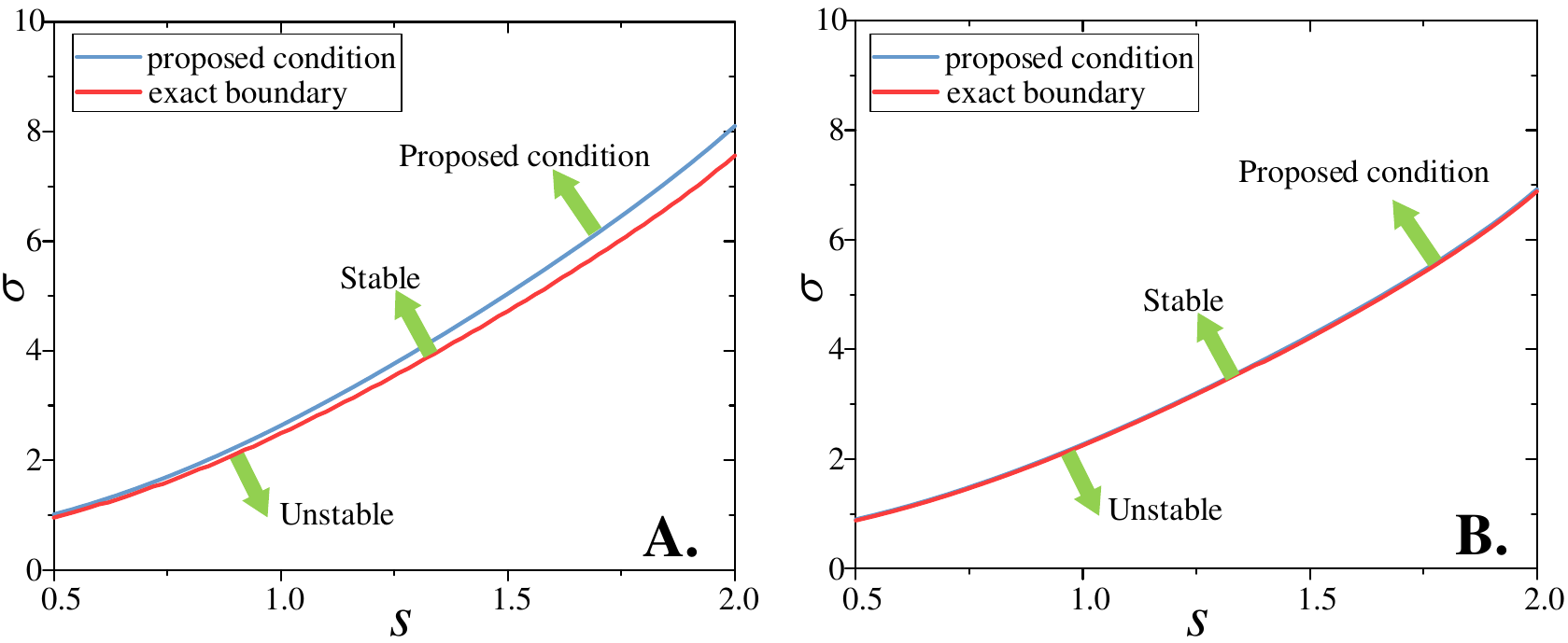}
	\caption{Stability verification of the IEEE 39-bus power system with (A) 24 bus dynamics and (B) 26 bus dynamics.}
	\label{fig:tight2}
\end{figure}

In order to demonstrate the scalability of our method, we now connect more dynamical components to the power system depicted in Fig.\ref{fig:39bus}. Firstly, we further equip the loads at buses 18 and 21 with a CD and a QD, respectively, such that the system possesses 24 heterogeneous bus dynamics in total. Secondly, buses 28 and 29 are also connected to a CD and a QD resulting in 26 dynamical buses. The stability verification results of these two augmented power systems are displayed in Fig.\ref{fig:tight2}. It is shown that no matter how many dynamical components are interconnected, the system-wide stability is always ensured as long as each participant obeys the proposed distributed conditions.

\section{Concluding Remarks}
In this paper, we derive distributed conditions for system-wide stability of lossy power systems with nonlinear heterogeneous bus dynamics. From the perspective of input-output interconnection, the system is divided into bus dynamics and the coupling strength. Leveraging the tailored concept of passivity index, we quantitatively evaluate the impact of each part and the assessment of system-wide stability is boiled down to a simple comparison between the coupling and the bus dynamics. Moreover, such an index can instruct proper controller designs, which is shown by three typical dynamical examples in the power systems. 

The proposed conditions offer a new approach to assess the system-wide stability in a distributed manner, shedding new light on the scalable stability analytic for the future power system with massive heterogeneous dynamic components.
In this regard, a crucial and attractive implication is that, it may allow constructing a certain "\emph{stability protocol}" to canonicalize the behaviors of a great variety of dynamic components connected to the power system, such that the system-wide stability can be well retained. 
An extension to the DAE power system model and controller designs for more general dynamics are our ongoing works. 

\section*{Acknowledgments}
The authors would like to thank Prof. Steven H. Low and Prof. Felix F. Wu for stimulating discussions.
\bibliographystyle{IEEEtran}
\bibliography{mybib}

\begin{thebibliography}{10}
\providecommand{\url}[1]{#1}
\csname url@samestyle\endcsname
\providecommand{\newblock}{\relax}
\providecommand{\bibinfo}[2]{#2}
\providecommand{\BIBentrySTDinterwordspacing}{\spaceskip=0pt\relax}
\providecommand{\BIBentryALTinterwordstretchfactor}{4}
\providecommand{\BIBentryALTinterwordspacing}{\spaceskip=\fontdimen2\font plus
\BIBentryALTinterwordstretchfactor\fontdimen3\font minus
  \fontdimen4\font\relax}
\providecommand{\BIBforeignlanguage}[2]{{%
\expandafter\ifx\csname l@#1\endcsname\relax
\typeout{** WARNING: IEEEtran.bst: No hyphenation pattern has been}%
\typeout{** loaded for the language `#1'. Using the pattern for}%
\typeout{** the default language instead.}%
\else
\language=\csname l@#1\endcsname
\fi
#2}}
\providecommand{\BIBdecl}{\relax}
\BIBdecl

\bibitem{Stott_Powersystemdynamic_1979}
B.~Stott, ``Power system dynamic response calculations,'' \emph{Proceedings of
  the IEEE}, vol.~67, no.~2, pp. 219--241, 1979.

\bibitem{Sastry_Hierarchicalstabilityalert_1980}
S.~Sastry and P.~Varaiya, ``Hierarchical stability and alert state steering
  control of interconnected power systems,'' \emph{IEEE Trans. Circuits Syst.},
  vol.~27, no.~11, pp. 1102--1112, 1980.

\bibitem{Chiang_Directstabilityanalysis_1995}
H.-D. Chang, C.-C. Chu, and G.~Cauley, ``Direct stability analysis of electric
  power systems using energy functions: theory, applications, and
  perspective,'' \emph{Proceedings of the IEEE}, vol.~83, no.~11, pp.
  1497--1529, 1995.

\bibitem{wang2018distributed_coping}
Z.~Wang, S.~Mei, F.~Liu, S.~H. Low, and P.~Yang, ``Distributed load-side
  control: Coping with variation of renewable generations,'' \emph{Automatica,
  in press}, 2019.

\bibitem{Wang2019Distributed_I}
Z.~Wang, F.~{Liu}, S.~H. {Low}, C.~{Zhao}, and S.~{Mei}, ``Distributed
  frequency control with operational constraints, part i: Per-node power
  balance,'' \emph{IEEE Trans. Smart Grid}, vol.~10, no.~1, pp. 40--52, Jan
  2019.

\bibitem{Wang2019Distributed_II}
Z.~{Wang}, F.~{Liu}, S.~H. {Low}, C.~{Zhao}, and S.~{Mei}, ``Distributed
  frequency control with operational constraints, part ii: Network power
  balance,'' \emph{IEEE Trans. Smart Grid}, vol.~10, no.~1, pp. 53--64, Jan
  2019.

\bibitem{MONSHIZADEH2019258}
N.~Monshizadeh and I.~Lestas, ``Secant and {{Popov}}-like conditions in power
  network stability,'' \emph{Automatica}, vol. 101, pp. 258 -- 268, 2019.

\bibitem{Song_DistributedFrameworkStability_2017}
Y.~Song, D.~J. Hill, T.~Liu, and Y.~Zheng, ``A distributed framework for
  stability evaluation and enhancement of inverter-based microgrids,''
  \emph{IEEE Trans. Smart Grid}, vol.~8, no.~6, pp. 3020--3034, 2017.

\bibitem{Zhang_OnlineDynamicSecurity_2015}
Y.~Zhang and L.~Xie, ``Online dynamic security assessment of microgrid
  interconnections in smart distribution systems,'' \emph{IEEE Trans. Power
  Syst.}, vol.~30, no.~6, pp. 3246--3254, 2015.

\bibitem{Ilic_standardsmodelbasedcontrol_2012}
M.~D. Ili{\'c} and Q.~Liu, ``Toward standards for model-based control of
  dynamic interactions in large electric power grids,'' in \emph{Proceedings of
  The 2012 Asia Pacific Signal and Information Processing Association Annual
  Summit and Conference}.\hskip 1em plus 0.5em minus 0.4em\relax IEEE, 2012,
  pp. 1--8.

\bibitem{Zhang_TransientStabilityAssessment_2016}
Y.~Zhang and L.~Xie, ``A transient stability assessment framework in power
  electronic-interfaced distribution systems,'' \emph{IEEE Trans. Power Syst.},
  vol.~31, no.~6, pp. 5106--5114, 2016.

\bibitem{Kundu_sumofsquaresapproachstability_2015}
S.~Kundu and M.~Anghel, ``A sum-of-squares approach to the stability and
  control of interconnected systems using vector lyapunov functions,'' in
  \emph{2015 American Control Conference (ACC)}.\hskip 1em plus 0.5em minus
  0.4em\relax IEEE, 2015, pp. 5022--5028.

\bibitem{Bao_ProcessControlPassive_2007}
J.~Bao and P.~L. Lee, \emph{Process control: the passive systems
  approach}.\hskip 1em plus 0.5em minus 0.4em\relax Springer Science \&
  Business Media, 2007.

\bibitem{vanderSchaft_L2GainPassivityTechniques_2017}
A.~J. van~der Schaft and A.~Van Der~Schaft, \emph{L2-gain and passivity
  techniques in nonlinear control}.\hskip 1em plus 0.5em minus 0.4em\relax
  Springer, 2000, vol.~2.

\bibitem{VanDerSchaft_PortHamiltoniansystemsnetwork_2004}
A.~Van~der Schaft, ``Port-hamiltonian systems: network modeling and control of
  nonlinear physical systems,'' in \emph{Advanced dynamics and control of
  structures and machines}.\hskip 1em plus 0.5em minus 0.4em\relax Springer,
  2004, pp. 127--167.

\bibitem{Fiaz_portHamiltonianapproachpower_2013a}
S.~Fiaz, D.~Zonetti, R.~Ortega, J.~M. Scherpen, and A.~Van~der Schaft, ``A
  port-hamiltonian approach to power network modeling and analysis,''
  \emph{European Journal of Control}, vol.~19, no.~6, pp. 477--485, 2013.

\bibitem{Caliskan_CompositionalTransientStability_2014}
S.~Y. Caliskan and P.~Tabuada, ``Compositional {{Transient Stability Analysis}}
  of {{Multimachine Power Networks}},'' \emph{IEEE Trans. Control Network
  Syst.}, vol.~1, no.~1, pp. 4--14.

\bibitem{8424071}
Z.~{Wang}, F.~{Liu}, J.~Z.~F. {Pang}, S.~H. {Low}, and S.~{Mei}, ``Distributed
  optimal frequency control considering a nonlinear network-preserving model,''
  \emph{IEEE Trans. Power Syst.}, vol.~34, no.~1, pp. 76--86, Jan 2019.

\bibitem{Stegink_unifyingenergybasedapproach_2016}
T.~Stegink, C.~De~Persis, and A.~van~der Schaft, ``A unifying energy-based
  approach to stability of power grids with market dynamics,'' \emph{IEEE
  Trans. Autom. Control}, vol.~62, no.~6, pp. 2612--2622, 2017.

\bibitem{trip2016internal}
S.~Trip, M.~B{\"u}rger, and C.~De~Persis, ``An internal model approach to
  (optimal) frequency regulation in power grids with time-varying voltages,''
  \emph{Automatica}, vol.~64, pp. 240--253, 2016.

\bibitem{Giusto_TransientStabilizationPower_2006}
A.~Giusto, R.~Ortega, and A.~Stankovic, ``On transient stabilization of power
  systems: A power-shaping solution for structure-preserving models,'' in
  \emph{Proceedings of the 45th IEEE Conference on Decision and Control}.\hskip
  1em plus 0.5em minus 0.4em\relax IEEE, 2006, pp. 4027--4031.

\bibitem{Sepulchre_ConstructiveNonlinearControl_1997}
R.~Sepulchre, M.~Jankovic, and P.~V. Kokotovic, \emph{Constructive nonlinear
  control}.\hskip 1em plus 0.5em minus 0.4em\relax Springer Science \& Business
  Media, 2012.

\bibitem{Li_ConsensusHeterogeneousMultiAgent_2019}
M.~Li, L.~Su, and G.~Chesi, ``Consensus of heterogeneous multi-agent systems
  with diffusive couplings via passivity indices,'' \emph{IEEE Control Syst.
  Lett.}, vol.~3, no.~2, pp. 434--439, 2019.

\bibitem{Yang_DistributedStabilityAnalytics_2019a}
P.~Yang, F.~Liu, Z.~Wang, C.~Shen, J.~Yi, and W.~Lin, ``Toward distributed
  stability analytics for power systems with heterogeneous bus dynamics,''
  \emph{arXiv preprint arXiv:1908.00752}, 2019.

\bibitem{Schiffer_surveymodelingmicrogrids_2016}
J.~Schiffer, D.~Zonetti, R.~Ortega, A.~M. Stankovi{\'c}, T.~Sezi, and
  J.~Raisch, ``A survey on modeling of microgrids—from fundamental physics to
  phasors and voltage sources,'' \emph{Automatica}, vol.~74, pp. 135--150,
  2016.

\bibitem{Simpson-Porco_VoltageStabilizationMicrogrids_2017}
J.~W. Simpson-Porco, F.~D{\"o}rfler, and F.~Bullo, ``Voltage stabilization in
  microgrids via quadratic droop control,'' \emph{IEEE Trans. Autom. Control},
  vol.~62, no.~3, pp. 1239--1253, 2017.

\bibitem{Kasis_PrimaryFrequencyRegulation_2017}
A.~Kasis, E.~Devane, C.~Spanias, and I.~Lestas, ``Primary frequency regulation
  with load-side participation—part i: Stability and optimality,'' \emph{IEEE
  Trans. Power Syst.}, vol.~32, no.~5, pp. 3505--3518, 2017.

\bibitem{Kundur_Powersystemstability_1994}
P.~Kundur, N.~J. Balu, and M.~G. Lauby, \emph{Power system stability and
  control}.\hskip 1em plus 0.5em minus 0.4em\relax McGraw-hill New York, 1994,
  vol.~7.

\bibitem{chiang1987foundations}
H.-D. Chiang, F.~Wu, and P.~Varaiya, ``Foundations of direct methods for power
  system transient stability analysis,'' \emph{IEEE Trans. Circuits Syst.},
  vol.~34, no.~2, pp. 160--173, 1987.

\bibitem{willems1972dissipative}
J.~C. Willems, ``Dissipative dynamical systems part i: General theory,''
  \emph{Archive for rational mechanics and analysis}, vol.~45, no.~5, pp.
  321--351, 1972.

\bibitem{Khalil_NonlinearSystems_2002}
H.~K. Khalil and J.~Grizzle, \emph{Nonlinear systems}.\hskip 1em plus 0.5em
  minus 0.4em\relax Prentice hall Upper Saddle River, NJ, 2002, vol.~3.

\bibitem{Jeltsema_Multidomainmodelingnonlinear_2009a}
D.~Jeltsema and J.~M. Scherpen, ``Multidomain modeling of nonlinear networks
  and systems,'' vol.~29, no.~4, pp. 28--59, 2009.

\bibitem{Yang_DistributedStabilityAnalytics_2019}
P.~Yang, F.~Liu, Z.~Wang, and S.~Ma, ``Towards distributed stability analytics
  of dynamic power systems: A phasor-circuit theory perspective,'' \emph{arXiv
  preprint arXiv:1907.12054}, 2019.

\bibitem{Ortega_Powershapingnew_2003}
R.~Ortega, D.~Jeltsema, and J.~M. Scherpen, ``Power shaping: A new paradigm for
  stabilization of nonlinear rlc circuits,'' \emph{IEEE Trans. Autom. Control},
  vol.~48, no.~10, pp. 1762--1767, 2003.

\bibitem{Kosaraju_KrasovskiiPassivity_2019}
K.~C. Kosaraju, Y.~Kawano, and J.~Scherpen, ``Krasovskii's passivity,''
  \emph{arXiv preprint arXiv:1903.05182}, 2019.

\bibitem{chiang2011direct}
H.-D. Chiang, \emph{Direct methods for stability analysis of electric power
  systems: theoretical foundation, BCU methodologies, and applications}.\hskip
  1em plus 0.5em minus 0.4em\relax John Wiley \& Sons, 2011.

\bibitem{Moon_Developmentenergyfunction_1997}
Y.~H. Moon, E.~H. Lee, and T.~H. Roh, ``Development of an energy function
  reflecting the transfer conductances for direct stability analysis in power
  systems,'' \emph{Transmission and Distribution IEE Proceedings - Generation},
  vol. 144, no.~5, pp. 503--509.

\bibitem{Ying_energybasedmethodologylocating_2012}
Y.~Li, C.~Shen, and F.~Liu, ``An energy-based methodology for locating the
  source of forced oscillations in power systems,'' in \emph{2012 IEEE
  International Conference on Power System Technology (POWERCON)}.\hskip 1em
  plus 0.5em minus 0.4em\relax IEEE, 2012, pp. 1--6.

\bibitem{chen2013energy}
L.~Chen, Y.~Min, and W.~Hu, ``An energy-based method for location of power
  system oscillation source,'' \emph{IEEE Trans. Power Syst.}, vol.~28, no.~2,
  pp. 828--836, 2013.

\bibitem{athay1979practical}
T.~Athay, R.~Podmore, and S.~Virmani, ``A practical method for the direct
  analysis of transient stability,'' \emph{IEEE Trans. Power Apparatus and
  Systems}, no.~2, pp. 573--584, 1979.

\bibitem{chiang2013line}
H.-D. Chiang, H.~Li, J.~Tong, and Y.~Tada, ``On-line transient stability
  screening of a practical 14,500-bus power system: Methodology and
  evaluations,'' in \emph{High Performance Computing in Power and Energy
  Systems}.\hskip 1em plus 0.5em minus 0.4em\relax Springer, 2013, pp.
  335--358.

\bibitem{Pogaku_ModelingAnalysisTesting_2007b}
N.~Pogaku, M.~Prodanovic, and T.~C. Green, ``Modeling, analysis and testing of
  autonomous operation of an inverter-based microgrid,'' \emph{IEEE Trans.
  Power Electron.}, vol.~22, no.~2, pp. 613--625, 2007.

\bibitem{schiffer2014conditions}
J.~Schiffer, R.~Ortega, A.~Astolfi, J.~Raisch, and T.~Sezi, ``Conditions for
  stability of droop-controlled inverter-based microgrids,'' \emph{Automatica},
  vol.~50, no.~10, pp. 2457--2469, 2014.

\bibitem{hiskens2013ieee}
I.~Hiskens, ``Ieee pes task force on benchmark systems for stability
  controls,'' \emph{Technical Report}, 2013.

\end{thebibliography}

\appendices
\makeatletter
\@addtoreset{equation}{section}
\@addtoreset{theorem}{section}
\makeatother

\section{Proof of Lemma \ref{le:1}}\label{app:A0}
\renewcommand{\theequation}{A.\arabic{equation}}
\renewcommand{\thetheorem}{A.\arabic{theorem}}
	\begin{proof}
		By the Taylor's theorem, there exists $\alpha\in[0,1]$, and $\psi=y^*+\alpha(y-y^*)$ such that $S_N(y)=S_N(y^*)+\nabla S_N(y^*)^T(y-y^*)+\frac{1}{2}(y-y^*)^T\nabla^2S_N(\psi)(y-y^*)$. From \eqref{eq:WB2}, we have $W_B(y^*)=0$. Hence, $S_N(y^*)=0$. Calculating the gradient of $S_N(y)$ yields $\nabla S_N(y)=\nabla \tilde{W}_B(y)-\nabla \tilde{W}_B(y^*)-\Sigma(y-y^*)$, where $\Sigma=\text{diag}\{(\sigma_1,\ldots,\sigma_n)\}$. Clearly we have $\nabla S_N(y^*)=0$. Hence, we have
		\begin{equation}\label{eq:midvalue}
		\setlength{\abovedisplayskip}{4pt}	
		\setlength{\belowdisplayskip}{4pt}
		S_N(y)=\frac{1}{2}(y-y^*)^T\nabla^2S_N(\psi)(y-y^*)
		\end{equation}
		Calculating the second-order partial derivatives (the Hessian matrix) of $S_N(y)$ yields
		\begin{equation*}
		\setlength{\abovedisplayskip}{4pt}	
		\setlength{\belowdisplayskip}{4pt}
		\nabla^2S_N(\psi)=\nabla^2\tilde{W}_B(\psi)+\Sigma \geq(\sigma+\lambda_{\min}(\nabla^2\tilde{W}_B(\psi)))I
		\end{equation*}
		According to the definition \eqref{eq:lambda} of $\lambda$, the condition $\sigma>-\lambda$ yields $\sigma>-\lambda_{\min}(\nabla^2\tilde{W}_B(y^*))$ and, hence, $\nabla^2S_N(y^*)>0$. Since $\tilde{W}_B(y)$ is a smooth function of $y$, entries of $\nabla^2\tilde{W}_B(y)$ are smooth in $y$ as well. Since eigenvalues of a matrix is continuous in its entries, by the continuity of composition, $\lambda_{\min}(\nabla^2\tilde{W}_B(y))$ is continuous in $y$. Therefore, there exists $\delta>0$ such that $\sigma>-\lambda_{\min}(\nabla^2\tilde{W}_B(y))$ holds $\forall y\in\mathcal{U}_\delta(y^*)$. It follows from \eqref{eq:midvalue} that $\forall y\in \mathcal{U}_\delta(y^*)\setminus\{y^*\}$ we have $S_N(y)>0$, which completes the proof.
	\end{proof}
	
\section{Proof of Theorem \ref{th:1}}\label{app:A}
\renewcommand{\theequation}{A.\arabic{equation}}
\renewcommand{\thetheorem}{A.\arabic{theorem}}
\begin{proof}
	1) Consider the following Lyapunov candidate 
	\begin{equation}\label{eq:Lyapunov}
	\setlength{\abovedisplayskip}{4pt}	
	\setlength{\belowdisplayskip}{4pt}
	W(x)=S_N(y)+\sum_{i\in\mathcal{V}}\big(S_i(x_i)-S_i(x_i^*)\big)
	\end{equation}
	where $S_N(y)$ and $S_i(x_i)$ are defined in \eqref{eq:SN_loss} and Condition \ref{c1}, respectively. 
	Since $\sigma_i>-\lambda$ holds $\forall i\in\mathcal{V}$, it follows from Lemma \ref{le:1} that $S_N(y)$ is locally positive-definite in some neighborhood of $y^*$. In addition, Condition \ref{c1} requires $\forall i\in\mathcal{V}$, $x_i^*$ is the local minimum of $S_i(x_i)$. Hence, it is guaranteed that there exists a neighborhood $\mathcal{U}_{\delta}(x^*)$ of $x^*$ such that $W(x^*)=0$ and $W(x)>0$, $\forall x\in\mathcal{U}_{\delta}(x^*)\setminus\{x^*\}$.

	Calculating the time derivative of $S_N(y)$ yields
	\begin{equation*}\label{eq:dotSN}
	\setlength{\abovedisplayskip}{4pt}	
	\setlength{\belowdisplayskip}{4pt}
	\begin{aligned}
	\dot{S}_N(y)=&(\nabla \tilde{W}_B(y)-\nabla \tilde{W}_B(y^*))^T\dot{y}+(y-y^*)^T\Sigma\dot{y}\\
	=&\sum_{i\in\mathcal{V}}\big[
	(P_i-P_i^*)\dot{\theta}_i+ (\frac{Q_i}{V_i}-\frac{Q_i^*}{V_i^*})\dot{V}_i+\sigma_i(y_i-y_i^*)^T\dot{y}_i\big]
	\end{aligned}
	\end{equation*}
	where  the first two terms of the second equality result from \eqref{eq:pfB}, and matrix $\Sigma:=\text{diag}\{(\sigma_1,\ldots,\sigma_n)\}$.
	
	From \eqref{eq:c1}, we have
	\begin{equation*}
	\setlength{\abovedisplayskip}{4pt}	
	\setlength{\belowdisplayskip}{4pt}
	\dot{W}(x)=\dot{S}_N(y)+\sum_{i\in\mathcal{V}}\dot{S}_i(x_i)\leq0
	\end{equation*}
	Thus, $W(x)$ is a (weak) Lyapunov function in $D:=\mathcal{U}_{\delta}(x^*)$ and $x^*$ is Lyapunov stable.
	
	2) By Condition \ref{c1'}, from \eqref{eq:c1'} we have
	\begin{equation}\label{eq:dotW}
	\setlength{\abovedisplayskip}{4pt}	
	\setlength{\belowdisplayskip}{4pt}
	\dot{W}(x)=\dot{S}_N(y)+\sum_{i\in\mathcal{V}}\dot{S}_i(x_i)\leq-\sum_{i\in\mathcal{V}}\varphi_i(\dot{y}_i)
	\end{equation}
	Since Condition \ref{c1'} requires $\varphi_i(\dot{y}_i)$ to be a positive-definite function, we have $\dot{W}\leq0$. Based on 1) $W(x)$ is a (weak) Lyapunov function in $D$. Hence, for a sufficiently small level value  $l$, there exists a compact set $\Xi:=\{x\in D:W(x)\leq l\}$, which is positively invariant of the interconnected system \eqref{eq:entire}. Since $x^*$ is an isolated equilibrium by Assumption \ref{as:isolate}, for a sufficiently small $l$, $\Xi$ contains no other equilibrium except $x^*$. By LaSalle's invariance principle \cite{Khalil_NonlinearSystems_2002}, any trajectory of system \eqref{eq:entire} starting with the finite initial value $x(0)\in\Xi$ converges to the largest invariant set of $\Upsilon:=\Xi\cap\{x:\dot{W}(x)=0\}$. It follows from \eqref{eq:dotW} and the positive-definiteness of $\varphi_i(\dot{y}_i)$ that $\dot{W}(x)=0$ implies $\dot{y}_i=0,\;\forall i\in\mathcal{V}$. By Condition \ref{c2}, $\dot{y}_i=0$, $\forall t\geq0$ indicates $\dot{x}_i=0$, $\forall t\geq0$. Therefore, the largest invariant set of $\Upsilon$ contains only the equilibrium points of system \eqref{eq:entire}. Thus, the trajectory converges to the unique equilibrium point $x^*$ in $\Upsilon$, which completes the proof. 
\end{proof}
\section{Proof of Theorem \ref{th:2}}\label{app:B}
\renewcommand{\theequation}{B.\arabic{equation}}
\renewcommand{\thetheorem}{B.\arabic{theorem}}
\begin{proof}
	1) Let the network storage function be 
	\begin{equation*}
	\setlength{\abovedisplayskip}{4pt}	
	\setlength{\belowdisplayskip}{4pt}
	\hat{S}_N(y):=S_N(y)+W_G(y)
	\end{equation*}
	where $S_N(y)$ and $W_G(y)$ are defined as \eqref{eq:SN_loss} and \eqref{eq:WG}, respectively. It follows from \eqref{eq:SN_loss} \eqref{eq:WB2} and \eqref{eq:WG} that $\hat{S}_N(y^*)=0$ and $\nabla\hat{S}_N(y^*)=0$. Calculating the Hessian matrix of $\hat{S}_N(y)$ yields 
	\begin{equation*}
	\setlength{\abovedisplayskip}{4pt}	
	\setlength{\belowdisplayskip}{4pt}
	\nabla^2\hat{S}_N(y)=\nabla^2\tilde{W}_B(y)+\frac{1}{2}\left(\nabla\Phi(y^*)+\nabla\Phi(y^*)^T\right)+\Sigma
	\end{equation*}
	where $\Sigma:=\text{diag}\{(\sigma_1,\ldots,\sigma_n)\}$.
	Following the lines in Lemma \ref{le:1}, one can prove that the condition $\sigma_i>-\lambda$ yields that there exists $\hat{\delta}>0$ such that $\nabla^2\hat{S}_N(y)>0$, $\forall y\in\mathcal{U}_{\hat{\delta}}(y^*)$ and consequently $\hat{S}_N(y)>0$, $\forall y\in \mathcal{U}_{\hat{\delta}}(y^*)\setminus\{y^*\}$.
	
	Now consider the following Lyapunov candidate 
	\begin{equation}\label{eq:Lyapunov2}
	\hat{W}(x)=\hat{S}_N(y)+\sum_{i\in\mathcal{V}}\big(S_i(x_i)-S_i(x_i^*)\big)
	\end{equation}
	Similarly, resulting from the positive-definiteness of $\hat{S}_N(y)$ and Condition \ref{c1}, there exists a neighborhood $\mathcal{U}_{\hat{\delta}}(x^*)$ of $x^*$ such that $\hat{W}(x^*)=0$ and $\hat{W}(x)>0$, $\forall x\in\mathcal{U}_{\hat{\delta}}(x^*)\setminus\{x^*\}$.
	
	Calculating the time derivative of $\hat{S}_N(y)$ leads to
	\begin{equation*}
	\setlength{\abovedisplayskip}{4pt}	
	\setlength{\belowdisplayskip}{4pt}
	\begin{aligned}
	\dot{\hat{S}}_N(y)=&(\nabla \tilde{W}_B(y)-\nabla \tilde{W}_B(y^*))^T\dot{y}+(\nabla W_G(y)-\nabla W_G(y^*))^T\dot{y}\\
	&+(y-y^*)^T\Sigma\dot{y}
	\end{aligned}
	\end{equation*}
	Bearing in mind \eqref{eq:pfsum} and the approximation \eqref{eq:approx}, we have
	\begin{equation*}
	\setlength{\abovedisplayskip}{4pt}	
	\setlength{\belowdisplayskip}{4pt}
	\dot{\hat{S}}_N(y)=\sum_{i\in\mathcal{V}}\big[
	(P_i-P_i^*)\dot{\theta}_i+ (\frac{Q_i}{V_i}-\frac{Q_i^*}{V_i^*})\dot{V}_i+\sigma_i(y_i-y_i^*)^T\dot{y}_i\big]
	\end{equation*}
	From \eqref{eq:c1}, we have
	\begin{equation*}
	\setlength{\abovedisplayskip}{4pt}	
	\setlength{\belowdisplayskip}{4pt}
	\dot{\hat{W}}(x)=\dot{\hat{S}}_N(y)+\sum_{i\in\mathcal{V}}\dot{S}_i(x_i)\leq0
	\end{equation*}
	Thus, $\hat{W}(x)$ is a (weak) Lyapunov function in $\hat{D}:=\mathcal{U}_{\hat{\delta}}(x^*)$ and $x^*$ is Lyapunov stable.
	
	2) By Condition \ref{c1'}, from \eqref{eq:c1'} we have
	\begin{equation}\label{eq:dotW2}
	\setlength{\abovedisplayskip}{4pt}	
	\setlength{\belowdisplayskip}{4pt}
	\dot{\hat{W}}(x)=\dot{\hat{S}}_N(y)+\sum_{i\in\mathcal{V}}\dot{S}_i(x_i)\leq-\sum_{i\in\mathcal{V}}\varphi_i(\dot{y}_i)
	\end{equation}
	Based on 1) $\hat{W}(x)$ is a (weak) Lyapunov function in $D$. Hence, for a sufficiently small level value  $l$, there exists a compact set $\Xi:=\{x\in D:\hat{W}(x)\leq l\}$, which is positively invariant of the interconnected system \eqref{eq:entire}. Since $x^*$ is an isolated equilibrium by Assumption \ref{as:isolate}, for a sufficiently small $l$, $\Xi$ contains no other equilibrium except $x^*$. By LaSalle's invariance principle \cite{Khalil_NonlinearSystems_2002}, any trajectory of system \eqref{eq:entire} starting with the finite initial value $x(0)\in\Xi$ converges to the largest invariant set of $\Upsilon:=\Xi\cap\{x:\dot{\hat{W}}(x)=0\}$. It follows from \eqref{eq:dotW2} and the positive-definiteness of $\varphi_i(\dot{y}_i)$ that $\dot{\hat{W}}(x_i)=0$ implies $\dot{y}_i=0,\;\forall i\in\mathcal{V}$. By Condition \ref{c2}, $\dot{y}_i=0$, $\forall t\geq0$ indicates $\dot{x}_i=0$, $\forall t\geq0$. Therefore, the largest invariant set of $\Upsilon$ contains only the equilibrium points of system \eqref{eq:entire}. Thus, the trajectory converges to the unique equilibrium point $x^*$ in $\Upsilon$, which completes the proof.  
\end{proof}

\section{Proof of Proposition \ref{pro:sg}}\label{app:C}
\renewcommand{\theequation}{C.\arabic{equation}}
\begin{proof}
	Consider the following storage function
	\begin{equation*}
	\setlength{\abovedisplayskip}{4pt}	
	\setlength{\belowdisplayskip}{4pt}
	\begin{aligned}
	S_i(x_i)=&\frac{1}{2}M_i\omega_i^2+\frac{K_I-\sigma_i}{2}(\delta_i-\delta_i^*)^2\\
	&+\frac{1}{2}(\frac{K_E+1}{x_{di}-x_{di}'}-\sigma_i)(E_{qi}'-E_{qi}'^*)^2
	\end{aligned}
	\end{equation*}
	Since $K_I>\sigma_i$ and $K_E>(x_{di}-x_{di}')\sigma_i-1$, it is clear that $x^*_i=(0,\delta_i^*,E_{qi}'^*)^T$ is the minimum of $S_i(x_i)$ and $S_i(x_i)>0$, $\forall x_i\neq x_i^*$.
	
	Calculating its time derivative, we have 
	\begin{equation}\label{eq:sgdotS}
	\setlength{\abovedisplayskip}{4pt}	
	\setlength{\belowdisplayskip}{4pt}
	\begin{aligned}
	\dot{S}_i=&M_i\omega_i\dot{\omega}_i+K_I(\delta_i-\delta_i^*)\dot{\delta}_i+\frac{K_E+1}{x_{di}-x_{di}'}(E_{qi}'-E_{qi}'^*)\dot{E}_{qi}'\\&-\sigma_i(y_i-y_i^*)^T\dot{y}_i
	\end{aligned}
	\end{equation}
	By \eqref{eq:sg1} and \eqref{eq:sgPg} we have
	\begin{equation}\label{eq:sgthe}
	\setlength{\abovedisplayskip}{4pt}	
	\setlength{\belowdisplayskip}{4pt}
	\begin{aligned}
	M_i\omega_i\dot{\omega}_i&=(-D_i\omega_i-P_i+P^g_i)\omega_i\\
	&=\big(-D_i\omega_i-P_i-K_I(\delta_i-\delta_i^*)-K_P\omega_i+P^*_i\big)\omega_i
	\end{aligned}
	\end{equation}
	Substituting \eqref{eq:sgEf} into \eqref{eq:sg1} and noticing the fact $-E_{qi}'^*-\frac{x_{di}-x_{di}'}{E_{qi}'^*}Q_i^*+E_{fi}^*=0$, we have
	\begin{equation*}
	\setlength{\abovedisplayskip}{4pt}	
	\setlength{\belowdisplayskip}{4pt}
	T_{di}'\dot{E}_{qi}'=-(K_E+1)(E_{qi}'-E_{qi}'^*)-(x_{di}-x_{di}')(\frac{Q_i}{E_{qi}'}-\frac{Q_i^*}{E_{qi}'^*})
	\end{equation*}
	which yields
	\begin{equation}\label{eq:sgV}
	\setlength{\abovedisplayskip}{4pt}	
	\setlength{\belowdisplayskip}{4pt}
	\frac{K_E+1}{x_{di}-x_{di}'}(E_{qi}'-E_{qi}'^*)=-(\frac{Q_i}{E_{qi}'}-\frac{Q_i^*}{E_{qi}'^*})-\frac{T_{di}'}{x_{di}-x_{di}'}\dot{E}_{qi}'
	\end{equation}
	Therefore, by substituting \eqref{eq:sgthe} and \eqref{eq:sgV} into \eqref{eq:sgdotS}, we have
	\begin{equation*}
	\setlength{\abovedisplayskip}{4pt}	
	\setlength{\belowdisplayskip}{4pt}
	\begin{aligned}
	\dot{S}_i=&-(P_i-P_i^*)\dot{\delta}_i-(\frac{Q_i}{E_{qi}'}-\frac{Q_i^*}{E_{qi}'^*})\dot{E}_{qi}'-\sigma_i(y_i-y_i^*)^T\dot{y}_i\\
	&-(D_i+K_P)\omega_i^2-\frac{T_{di}'}{x_{di}-x_{di}'}(\dot{E}_{qi}')^2
	\end{aligned}
	\end{equation*}
	Since $K_P>-D_i$ and $x_{di}-x_{di}'>0$, Condition \ref{c1'} is satisfied with $\varphi_i(\dot{y}_i)=(D_i+K_P)\dot{\delta}_i^2+\frac{T_{di}'}{x_{di}-x_{di}'}(\dot{E}_{qi}')^2$.
	
	If $\dot{\delta}_i=\omega_i=0$, $\forall t\geq0$, then $\dot{\omega}_i=0$, $\forall t\geq0$ holds as well. Thus, $\dot{y}_i=0$, $\forall t\geq0$ implies $\dot{x}_i=0$, $\forall t\geq0$, which means Condition \ref{c2} is also satisfied.
\end{proof}
\section{Proof of Proposition \ref{pro:CD}}\label{app:D}
\renewcommand{\theequation}{D.\arabic{equation}}
\begin{proof}
	Consider the following storage function
	\begin{equation*}
	\setlength{\abovedisplayskip}{4pt}	
	\setlength{\belowdisplayskip}{4pt}
	S_i(x_i)=\frac{D_{i1}^{-1}-\sigma_i}{2}(\theta_i-\theta_i^*)^2+\frac{k_i}{D_{i2}}(\frac{V_i}{V_i^*}-\ln V_i)-\frac{\sigma_i(V_i-V_i^*)^2}{2}
	\end{equation*}
	where $k_i=V_i^*+D_{i2}Q_i^*$ is a constant.
	Calculating the gradient of $S_i(x_i)$ yields
	\begin{equation*}
	\setlength{\abovedisplayskip}{4pt}	
	\setlength{\belowdisplayskip}{4pt}
	\nabla S_i(x_i)=(D_{i1}^{-1}-\sigma_i)(\theta_i-\theta_i^*)+\frac{k_i}{D_{i2}}(\frac{1}{V_i^*}-\frac{1}{V_i})-\sigma_i(V_i-V_i^*)
	\end{equation*}
	And its Hessian matrix is
	\begin{equation*}
	\setlength{\abovedisplayskip}{4pt}	
	\setlength{\belowdisplayskip}{4pt}
	\nabla^2S_i(x_i)=\text{diag}\{(D_{i1}^{-1}-\sigma_i,\frac{k_i}{D_{i2}V_i^2}-\sigma_i)\}
	\end{equation*}
	Clearly, we have $\nabla S_i(x_i^*)=0$. 
	Bearing in mind $V_i>0$ and $k_i=V_i^*+D_{i2}Q_i^*$, we have
	\begin{equation*}
	\setlength{\abovedisplayskip}{4pt}	
	\setlength{\belowdisplayskip}{4pt}
	\frac{k_i}{D_{i2}V_i^{*2}}-\sigma_i>0\Leftrightarrow\frac{V_i^*}{D_{i2}}>\sigma_iV_i^{*2}-Q_i^*\Leftrightarrow \frac{1}{D_{i2}}>\frac{V_i^{*2}\sigma_i-Q_i^*}{V_i^*}
	\end{equation*}
	Thus, \eqref{eq:conventionD} yields $\nabla^2S_i(x_i^*)>0$. Therefore, $x_i^*=(\theta_i^*,V_i^*)^T$ is a strict local minimum of $S_i(x_i)$.
	
	In addition, calculating its time derivative yields 
	\begin{equation}\label{eq:CDdotS}
	\setlength{\abovedisplayskip}{4pt}	
	\setlength{\belowdisplayskip}{4pt}
	\dot{S}_i=D_{i1}^{-1}(\theta_{i}-\theta_{i}^*)\dot{\theta}_i+\frac{k_i}{D_{i2}}(\frac{1}{V_i^*}-\frac{1}{V_i})\dot{V}_i-\sigma_i(y_i-y_i^*)^T\dot{y}_i
	\end{equation}
	By the first equation of \eqref{eq:egV}, we have
	\begin{equation}\label{eq:CDthe}
	\setlength{\abovedisplayskip}{4pt}	
	\setlength{\belowdisplayskip}{4pt}
	\begin{aligned}
	D_{i1}^{-1}(\theta_{i}-\theta_{i}^*)&=D_{i1}^{-1}(-\tau_{i1}\dot{\theta}_i-D_{i1}(P_i-P_i^*))\\
	&=-(P_i-P_i^*)-\frac{\tau_{i1}}{D_{i1}}\dot{\theta}_i
	\end{aligned}
	\end{equation}
	Dividing the second equation of \eqref{eq:egV} by $V_i$, yields
	\begin{equation*}
	\setlength{\abovedisplayskip}{4pt}	
	\setlength{\belowdisplayskip}{4pt}
	\begin{aligned}
	\tau_{i2}\frac{\dot{V}_i}{V_i}&=-(1-\frac{V_i^*}{V_i})-D_{i2}(\frac{Q_i}{V_i}-\frac{Q_i^*}{V_i^*}+\frac{Q_i^*}{V_i^*}-\frac{Q_i^*}{V_i})\\
	&=-V_i^*(\frac{1}{V_i^*}-\frac{1}{V_i})-D_{i2}(\frac{Q_i}{V_i}-\frac{Q_i^*}{V_i^*})-D_{i2}Q_i^*(\frac{1}{V_i^*}-\frac{1}{V_i})\\
	&=-k_i(\frac{1}{V_i^*}-\frac{1}{V_i})-D_{i2}(\frac{Q_i}{V_i}-\frac{Q_i^*}{V_i^*})
	\end{aligned}
	\end{equation*}
	which leads to
	\begin{equation}\label{eq:CDv}
	\setlength{\abovedisplayskip}{4pt}	
	\setlength{\belowdisplayskip}{4pt}
	\frac{k_i}{D_{i2}}(\frac{1}{V_i^*}-\frac{1}{V_i})=-(\frac{Q_i}{V_{i}}-\frac{Q_i^*}{V_{i}^*})-\frac{\tau_{i2}}{D_{i2}V_i}\dot{V}_i
	\end{equation}
	Substituting \eqref{eq:CDthe} and \eqref{eq:CDv} into \eqref{eq:CDdotS} yields
	\begin{equation}
	\setlength{\abovedisplayskip}{4pt}	
	\setlength{\belowdisplayskip}{4pt}
	\begin{aligned}
	\dot{S}_i=&-(P_i-P_i^*)\dot{\theta}_i-(\frac{Q_i}{V_{i}}-\frac{Q_i^*}{V_{i}^*})\dot{V}_{i}-\sigma_i(y_i-y_i^*)^T\dot{y}_i\\
	&-\frac{\tau_{i1}}{D_{i1}}\dot{\theta}_i^2-\frac{\tau_{i2}}{D_{i2}V_i}\dot{V}_i^2
	\end{aligned}
	\end{equation}
	which meets Condition \ref{c1'}.
	
	Since in this case $x_i=y_i=\text{col}(\theta_{i},V_i)$, Condition \ref{c2} is trivially satisfied, which completes the proof.
\end{proof}

\section{Proof of Proposition \ref{pro:QD}}\label{app:E}
\renewcommand{\theequation}{E.\arabic{equation}}
\begin{proof}
	Consider the following storage function 
	\begin{equation*}
	\setlength{\abovedisplayskip}{4pt}	
	\setlength{\belowdisplayskip}{4pt}
	S_i(x_i)=\frac{D_{i1}^{-1}-\sigma_i}{2}(\theta_i-\theta_i^*)^2+\frac{D_{i2}^{-1}-\sigma_i}{2}(V_i-V_i^*)^2
	\end{equation*}
	The parameter range \eqref{eq:QDroop} yields that $x_i^*=(\theta_i^*,V_i^*)^T$ is a minimum of $S_i(x_i)$ and $S_i(x_i)>0$, $\forall x_i\neq x_i^*$.
	Calculating its time derivative, we have
	\begin{equation}\label{eq:QDdotS}
	\setlength{\abovedisplayskip}{4pt}	
	\setlength{\belowdisplayskip}{4pt}
	\dot{S}_i=D_{i1}^{-1}(\theta_{i}-\theta_{i}^*)\dot{\theta}_i+D_{i2}^{-1}(V_{i}-V_{i}^*)\dot{V}_i-\sigma_i(y_i-y_i^*)^T\dot{y}_i
	\end{equation}
	By \eqref{eq:egQD}, we have
	\begin{equation}\label{eq:QDthe}
	\setlength{\abovedisplayskip}{4pt}	
	\setlength{\belowdisplayskip}{4pt}
	D_{i1}^{-1}(\theta_{i}-\theta_{i}^*)=-(P_i-P_i^*)-\frac{\tau_{i1}}{D_{i1}}\dot{\theta}_i
	\end{equation}
	and
	\begin{equation*}
	\setlength{\abovedisplayskip}{4pt}	
	\setlength{\belowdisplayskip}{4pt}
	\begin{aligned}
	\tau_{i2}\frac{\dot{V}_i}{V_i}=-D_{i2}\frac{Q_i}{V_i}-(V_i-u_i^*)
	=-D_{i2}(\frac{Q_i}{V_i}-\frac{Q_i^*}{V_i^*})-(V_i-V_i^*)
	\end{aligned}
	\end{equation*}
	where the second equality is obtained by eliminating $u_i^*$ according to \eqref{eq:QDu}. Consequently, this leads to
	\begin{equation}\label{eq:QDv}
	\setlength{\abovedisplayskip}{4pt}	
	\setlength{\belowdisplayskip}{4pt}
	D_{i2}^{-1}(V_{i}-V_{i}^*)=-(\frac{Q_i}{V_{i}}-\frac{Q_i^*}{V_{i}^*})-\frac{\tau_{i2}}{D_{i2}V_i}\dot{V}_i
	\end{equation}
	Substituting \eqref{eq:QDthe} and \eqref{eq:QDv} into \eqref{eq:QDdotS} yields
	\begin{equation}
	\setlength{\abovedisplayskip}{4pt}	
	\setlength{\belowdisplayskip}{4pt}
	\begin{aligned}
	\dot{S}_i=&-(P_i-P_i^*)\dot{\theta}_i-(\frac{Q_i}{V_{i}}-\frac{Q_i^*}{V_{i}^*})\dot{V}_{i}-\sigma_i(y_i-y_i^*)^T\dot{y}_i\\
	&-\frac{\tau_{i1}}{D_{i1}}\dot{\theta}_i^2-\frac{\tau_{i2}}{D_{i2}V_i}\dot{V}_i^2
	\end{aligned}
	\end{equation}
	which meets Condition \ref{c1'}.
	
	Since in this case $x_i=y_i=\text{col}(\theta_{i},V_i)$, Condition \ref{c2} is trivially satisfied, which completes the proof.
\end{proof}

\end{document}